\documentclass[aps,12pt,amsfonts,amsmath,amssymb,nofootinbib]{revtex4}
\begin{document}
\newcommand{\pic}{$\spadesuit$}
\newcommand{\N}{\mathbb{N}}
\newcommand{\Z}{\mathbb{Z}}
\newcommand{\C}{\mathbb{C}}
\newcommand{\R}{\mathbb{R}}
\newcommand{\F}{\mathbb{F}}
\newcommand{\defineL}{\,\mathrel{\mathop:}=}
\newcommand{\defineR}{ =\mathrel{\mathop:}\,}
\newcommand{\be}{\begin{equation}}
\newcommand{\ee}{\end{equation}}
\newcommand{\bi}{\begin{itemize}}
\newcommand{\ei}{\end{itemize}}
\newcommand{\with}{\qquad\text{with}\qquad}
\newcommand{\mand}{\qquad\text{and}\qquad}
\newcommand{\sep}{\; \text{,}\qquad}
\newcommand{\ssep}{\: \text{,}\quad}
\newcommand{\pd}{\partial}
\newcommand{\pdo}{\overline{\partial}}
\newcommand{\ov}[1]{\overline{#1}}
\newcommand{\inv}[1]{\frac{1}{#1}}
\newcommand{\tinv}[1]{\tfrac{1}{#1}}
\newcommand{\abs}[1]{|#1|}
\newcommand{\B}[1]{\mathbf{#1}}
\newcommand{\op}[1]{\text{#1}}
\newcommand{\bracket}[1]{\left( #1 \right)}
\newcommand{\bracketi}[1]{\bigl( #1 \bigr)}
\newcommand{\bracketii}[1]{\Bigl( #1 \Bigr)}
\newcommand{\bracketiii}[1]{\biggl( #1 \biggr)}
\newcommand{\bracketiv}[1]{\Biggl( #1 \Biggr)}
\newcommand{\bpm}{\begin{pmatrix}}
\newcommand{\epm}{\end{pmatrix}}
\newcommand{\w}{\wedge}
\newcommand{\tp}{\otimes}
\newcommand{\ds}{\oplus}
\newcommand{\no}{\nonumber}
\newcommand{\DD}{\mathfrak{D}}
\newcommand{\LL}{\mathfrak{L}}
\newcommand{\FF}{\mathfrak{F}}
\newcommand{\cF}{\ti{\mathcal{F}}}
\newcommand{\cM}{\ti{\mathcal{M}}}
\newcommand{\cN}{\ti{\mathcal{N}}}
\newcommand{\cc}{\mathfrak{c}}
\newcommand{\coder}{\mathfrak{coder}}
\newcommand{\morph}{\mathfrak{morph}}
\newcommand{\hati}[1]{\hat{\tilde{#1}}}
\newcommand{\ti}[1]{\tilde{#1}}
\newcommand{\cp}{{c^\prime}}

\title{Quantum Open-Closed Homotopy Algebra\\ and String Field Theory\vspace{1.0cm}}

\author{Korbinian M\"unster}
\email{korbinian.muenster@physik.uni-muenchen.de} 
\affiliation{Arnold Sommerfeld Center for Theoretical Physics, Theresienstrasse 37, D-80333 Munich, Germany}
\author{Ivo Sachs}
\email{sachs@physics.harvard.edu}
\affiliation{Center for the Fundamental Laws of Nature, Harvard University, Cambridge, MA 02138, USA\\
and\\
Arnold Sommerfeld Center for Theoretical Physics, Theresienstrasse 37, D-80333 Munich, Germany}
\date{\today}

\begin{abstract}
\vspace{1.7cm}
\centerline{\bf Abstract \vspace{0.4cm}}
We reformulate the algebraic structure of Zwiebach's quantum open-closed string field theory in terms of homotopy algebras. We call it the quantum open-closed homotopy algebra (QOCHA) which is the generalization of the open-closed homotopy algebra (OCHA) of Kajiura and Stasheff. The homotopy formulation reveals new insights about deformations of open string field theory by closed string backgrounds. In particular, deformations by  Maurer Cartan elements of the quantum closed homotopy algebra define consistent quantum open string field theories.
\end{abstract}

\maketitle
\thispagestyle{empty}
\newpage
\tableofcontents
\newpage


\section{Introduction}\label{sec:intro}
String field theory is an off-shell formulation of string theory. Such a description is probably indispensable for a more fundamental understanding of string theory,  in particular, its underlying symmetries and the relation between open and closed strings (e.g. \cite{Kapustin:2004df}). On the other hand,  string field theory allows to address  non-perturbative phenomena such as tachyon condensation for instance (see e.g. \cite{Fuchs tachyoncond} for a review and references). 

The problem of constructing a string field theory is essentially that of the decomposition of moduli spaces $\mathcal{P}$ of bordered Riemann surfaces with closed string insertions in the bulk and open string insertions on the boundaries \cite{Zwiebach open-closed}. The most efficient way of doing so is based on the 
Batalin-Vilkovisky (BV) formalism, the simplest realization being  Witten's cubic, bosonic, open string field theory \cite{Witten:1986qs}. This theory realizes a differential graded algebra (DGA), with the differential given by the open string BRST operator. More generally, the vertices of any consistent  classical open string field theory satisfy the relations of an $A_\infty$-algebra (strongly homotopy associative algebra) \cite{Zwiebach open1}, which is a non-associative generalization of a DGA. In closed string theory there is no decomposition of the moduli space of Riemann surfaces compatible with Feynman rules obtained from a  cubic action. Consequently one has to introduce higher string vertices and as a result the closed string field theory becomes non-polynomial. For the same reason the algebraic structure takes the form of a $L_\infty$-algebra \cite{Zwiebach closed}, that is, a  differential graded Lie algebra up to homotopy. 

Consider now classical open-closed string field theory\footnote{This corresponds to taking the limit $\hbar\to 0$ after absorbing $\hbar^{\frac{1}{2}}$ in the closed string field. In this normalization the closed string anti-bracket is proportional to $\hbar$.}. This means that we include, in addition to the open string vertices with insertions on the boundary of the disc, and the closed string vertices with insertions on the sphere, disc vertices with an arbitrary number of open and closed string insertions. The set of these vertices satisfies the classical BV equation of open and closed strings to  $0$-th order in $\hbar$. The point of the reformulation of this in terms of homotopy algebras is that it reveals a new structure that is not explicit in the BV formulation:  The $A_\infty$-structure of a consistent open string theory  endows the space of generic vertices, with a (Hochschild) differential, $d_h$. This differential, together with the Gerstenhaber bracket $[\cdot,\cdot]$ in turn, imply the structure of a differential graded Lie algebra (DGL). Now, an useful insight of Kajiura and Stasheff \cite{Kajiura open-closed,Kajiura open-closed2} was that the disc vertices with open and closed inputs can be interpreted as a $L_\infty$-morphism from the $L_\infty$-algebra of closed strings to the DGL on the cyclic Hochschild complex of open string vertices. That defines the open-closed homotopy algebra (OCHA). Now, an important property of $L_\infty$-morphisms is that they map Maurer Cartan elements into Maurer Cartan elements. Maurer Cartan elements on the closed string side represent solutions of the equations of motion of closed string field theory - classical closed string backgrounds. On the other hand, Maurer Cartan elements on the open string side define a consistent, classical open string field theory.  Thus the $L_\infty$-morphism realized by the open-closed  vertices on the disc associates to every classical closed string background a consistent classical open string field theory.


%
%

In this paper we generalize the OCHA to the quantum level. That is we do not restrict to vertices with genus zero and at most one boundary, but include all vertices with arbitrary genus and arbitrary number of boundaries. Consequently we have to consider the full quantum BV master equation, which involves besides the odd Poisson bracket (antibracket) also the BV operator. The algebraic structure of quantum closed string field theory reformulated in homotopy language is called loop algebra \cite{Markl closed}. This is a special case of a more general algebraic structure, namely an involutive Lie bialgebra up to homotopy ($IBL_\infty$-algebra) \cite{Cieliebak ibl}. Furthermore, it has been realized recently that the cyclic Hochschild complex is equipped with a richer structure than just a Lie algebra, one can define an involutive Lie bialgebra (IBL-algebra) on it \cite{Chen liebi, Cieliebak ibl}. The main result of this paper is that the algebraic structure of quantum open-closed string field theory can be described by an $IBL_\infty$-morphism form the loop algebra of closed strings to the $IBL$-algebra defined on the cyclic Hochschild complex of open strings. 

The property that Maurer Cartan elements are mapped into Maurer Cartan elements holds also for $IBL_\infty$-morphisms. The $IBL_\infty$-morphism thus maps Maurer Cartan elements of the quantum closed string theory into a consistent quantum theories with only open strings. This is the quantum version of the open-closed correspondence (QOCHA). 
On the other hand, we show that the quantum closed string Maurer Cartan equation implies that the closed string BRST operator on the corresponding classical closed string background has to have trivial cohomology. This is in agreement what is known about the inconsistency of open string filed theory due to the presence of the closed string tadpole. 

The paper is organized as follows. In section \ref{summary} we  give a concise description of the concepts involved and summarize the main results. In section \ref{sai} we introduce $A_\infty$ and $L_\infty$ algebras. The material in this section is standard. It is nevertheless included to make the paper self-contained and accessible to mathematicians as well as physicists. In section \ref{hlb} $IBL_\infty$ algebras are introduced as a generalization of $L_\infty$ algebras.  In section \ref{sqocLBI} we explain how $A_\infty$/$L_\infty$ and, in particular, $IBL_\infty$ algebras are realized in open-closed string theory. The main result of this section is the realization of quantum open-closed string theory as an $IBL_\infty$ homotopy algebra with the open-closed vertices realizing an $IBL_\infty$-morphism. This is the advertised quantum generalization 
of the open-closed homotopy algebra (OCHA) of Kajiura and Stasheff. In section \ref{sec:defn} we analyze the closed string Maurer Cartan equation and its relation to consistent quantum open string field theory. In particular,  we show that the closed string Maurer Cartan equation implies that the closed string BRST cohomology in the corresponding classical closed string background is trivial. Appendix A contains a short description of the symplectic structure in open-closed string field theory. The detailed proof of the equivalence of the quantum open-closed BV-equation and the quantum open-closed homotopy algebra is contained in appendix B.

\section{Summary}\label{summary}
Since this paper is rather technical we will start with a summary of the main results leaving the technical details and most definitions to the later sections. 
 Let $A_o$ and $A_c$ denote the space of open and closed string fields respectively. These spaces are equipped with a grading - the ghost number. The quantum BV action $S$, of Zwiebach's open-closed string field theory \cite{Zwiebach open-closed}, is a collection of vertices with an arbitrary number of open and closed insertions, an arbitrary number of boundaries and arbitrary genus. Each vertex is invariant under the following transformations:
\bi
\item[(i)]cyclic permutation of open string inputs of one boundary
\item[(ii)]arbitrary permutation of closed string inputs
\item[(iii)]arbitrary permutation of boundaries
\ei
Let $\mathcal{V}^{b,g}_{n,m_1,\dots,m_b}$ denote the vertex of genus $g$ with $n$ closed string insertions and $b$ boundaries with $m_i$ representing the number of insertions on the $i$-th boundary.
This vertex comes with a certain power in $\hbar$ which is $2g+b+n/2-1$ \cite{Zwiebach open-closed}. The full BV action reads
\be\no
S(c,a)=\sum_{b,g}\sum_{n}\sum_{m_1,\dots,m_b}\hbar^{2g+b+n/2-1}\;\mathcal{V}^{b,g}_{n,m_1,\dots,m_b}(c,a)\;\text{,}
\ee 
where $c\in A_c$ is the closed string field and $a\in A_o$ is the open string field. $A_o$ and $A_c$ are modules over some ring $R$. In order to define a consistent quantum theory, the action $S$ has to satisfy the quantum BV master equation 
\be\label{eq:BVeqsummary}
\hbar \Delta^{BV} S+\inv{2}(S,S)=0\;\text{,}
\ee
where $\Delta^{BV}$ denotes the BV operator and $(\cdot,\cdot)$ denotes the odd Poisson bracket also known as antibracket \cite{Zwiebach open-closed}. These operations are constructed with the aid of the odd symplectic structures $\omega_o$ and $\omega_c$ on $A_o$ and $A_c$ respectively. The BV equation (\ref{eq:BVeqsummary}) puts constraints on the collection of vertices 
$\mathcal{V}^{b,g}_{n,m_1,\dots,m_b}$ and our goal is to interpret these constraints in the language of homotopy algebras.

The idea is to split the set of all vertices into two disjoint sets. One contains all vertices with closed string insertions only and the other contains all vertices with at least one open string input. Let us focus on the set 
of vertices with only closed string insertions first. Since $\omega_c$ is non-degenerate and the vertices are invariant w.r.t. any permutation of the inputs there is a unique map 
$l^g\in\op{Hom}^{cycl}(SA_c,A_c)$ such that 
\be\no
\mathcal{V}^{0,g}_{n}(c)=\inv{n!}\omega_c\bracket{l^g_{n-1}(c^{\w n-1}),c}  \;\text{,}\quad \forall g \;\text{.}
\ee
with $c^{\wedge n}\in SA_c$, the graded symmetric algebra of $A_c$. 
Upon summing over $n$ we then write   
\begin{align}\no
\sum_{n}\hbar^{2g+n/2-1}\;\mathcal{V}^{0,g}_{n}(c)&= \hbar^{2g-1}(\omega_c \circ l^g)(e^{\hbar^{1/2}c})\;\text{,}
\end{align}
where $\omega_c$ is interpreted as a map from $A_c$ to $A_c^\ast$. Following the same reasoning and taking all symmetries of vertices with open and closed inputs into account we can write
\be\label{ocv}
\sum_{n}\sum_{m_1,\dots,m_b}\hbar^{2g+b+n/2-1}\;\mathcal{V}^{b,g}_{n,m_1,\dots,m_b}(c,a)
=\inv{b!}\hbar^{2g+b-1} ({\omega}_o^{\tp b}\circ {f}^{b,g})(e^{\hbar^{1/2}c};\underbrace{\bar{e}^a,\dots,\bar{e}^a}_{\text{$b$ times}})\;\text{,}
\ee
where ${f}^{b,g}\in \op{Hom}(SA_c,R)\tp (\op{Hom}^{cycl}(TA_o,A_o))^{\w b}$. Furthermore, $\bar{e}^a\defineL\sum_{n=1}^\infty \inv{n}a^{\tp n}$ and $TA_o$ denotes the tensor algebra of $A_o$.  To summarize, the full BV quantum action of open-closed string field theory can be expressed as
\be\label{eq:BVactionsummary}
S=\sum_{g=0}^\infty \hbar^{2g-1}(\omega_c\circ l^g)(e^{\hbar^{1/2}c})
+ \sum_{b=1}^\infty\sum_{g=0}^\infty \inv{b!}\hbar^{2g+b-1}({\omega}^{\tp b}_o\circ {f}^{b,g})(e^{\hbar^{1/2}c};\underbrace{\bar{e}^a,\dots,\bar{e}^a}_{\text{$b$ times}})\;\text{.}
\ee
The classical  open-closed homotopy algebra is then realized as follows: The maps of genus zero, $l^0$, parametrizing the classical closed string vertices (spheres) in (\ref{eq:BVactionsummary}) define a $L_\infty$-algebra, that is, there is a coderivation, $L_{cl}: SA_c\to SA_c$ with $ L_{cl}^2=0$. Similarly, the vertices of any consistent classical open string field theory realize an $A_\infty$-algebra defined by a coderivation, $M_{cl}: TA_o\to TA_o,\; M_{cl}^2=0$. This makes the space $\op{Coder}^{cycl}(TA_o)$ of  coderivations on $TA_o$, a differential graded Lie algebra, where $[\cdot,\cdot]$ is simply defined by the graded commutator of coderivations and $d_h\defineL[M_{cl},\cdot]$. This a special case of a $L_\infty$-algebra. The set of open-closed disc vertices parametrized by ${f}^{1,0}$ is then identified as a $L_\infty$-morphism between the $L_\infty$-algebra of closed strings and the DGL on the cyclic Hochschild complex of open strings vertices
\be\label{eq:OCHAformsummary}
(A_c,L_{cl})\xrightarrow[]{L_\infty-\text{morphism}} (\op{Coder}^{cycl}(TA_o),d_h,[\cdot,\cdot])\;\text{,}\nonumber
\ee
This is the open-closed homotopy algebra of Kajiura and Stasheff. 

We shall be interested in the quantum version of this homotopy algebra.  This  works as follows:  The closed string BV operator $\Delta^{BV}$ requires the inclusion of a so-called second order coderivation, $D(\omega_c^{-1})\in \op{Coder}^2(SA_c)$ defined by
\be
\pi_1\circ D(\omega_c^{-1})=0 \mand \pi_2 \circ D(\omega_c^{-1})=\omega_c^{-1}\;\text{,}\nonumber
\ee
that is, $D(\omega_c^{-1})$ has no inputs but two outputs. On the other hand, composition of a first order coderivation with $D(\omega_c^{-1})$  gives again a coderivation where two inputs have been glued together. In this way one produces new objects, $L^g\in\op{Coder}^{cycl}(SA_c)$, again equivalent to maps $l^g\in \op{Hom}^{cycl}(SA_c,A_c)$ which, in turn, represent closed string vertices corresponding to Riemann surfaces of higher genus.  The combination
\be\no
\LL_c= \sum_{g=0}^\infty \hbar^g L^g +  \hbar  D(\omega_c^{-1})\nonumber
\ee
together with the condition $\LL_c^2=0$ defines the homotopy loop algebra of closed string field theory  \cite{Markl closed}. This is a special case of an $IBL_\infty$-algebra.

We have already mentioned that the space of open string vertices, described by cyclic coderivations $\op{Coder}^{cycl}(TA_o)$ form a Lie algebra. 
However, it turns out that we can make $\op{Coder}^{cycl}(TA_o)$ even a differential involutive Lie bialgebra, i.e. there is a map $\delta:\op{Coder}^{cycl}(TA_o)\to \op{Coder}^{cycl}(TA_o)^{\tp 2}$ such that $[\cdot,\cdot]$ and $\delta$ satisfy the defining equations of an IBL-algebra. Concretely we define
\be\no
{\LL}_o\defineL {[\cdot,\cdot]}+\hbar {\delta}\nonumber
\ee
which satisfies ${\LL}_o^2=0$. This then allows us to  define the quantum open-closed homotopy algebra (QOCHA) as an $IBL_\infty$-morphisms from the $IBL_\infty$-algebra of closed strings to the $IBL$-algebra of open strings
\be
(A_c,\LL_c)\xrightarrow[]{IBL_\infty-\text{morphism}} (\op{Coder}^{cycl}(TA_o),{\LL}_o)\;\text{,}\nonumber
\ee
with
\be\no
{\FF}\circ\LL_c={\LL}_o\circ {\FF} .\nonumber
\ee
The $IBL_\infty$-morphism, $\FF$ is determined by the open-closed vertices $f^{b,g}$.  This is the main mathematical result of this paper. 

In order to get a grasp of the usefulness of QOCHA we now focus on the Maurer Cartan elements in homotopy algebras. Consider a purely open string theory with vertices described by some coderivation  ${\cal{M}}$. Quantum consistency of purely open string field theory then requires that ${\LL}_o(e^{\cal{M}})=0$. This is just the BV equation (\ref{eq:BVeqsummary}). Since  $IBL_\infty$-morphisms map Maurer Cartan elements into Maurer Cartan elements we can look for ${\cal{M}}$ in the image of ${\FF}$. In this way we are guaranteed to find a consistent open string field theory if 
\be\no
e^{\cal{M}}={\FF}(e^{\mathfrak{c}})
\ee
for some Maurer Cartan element, ${\LL}_c(e^{\mathfrak{c}})=0$. In order to see what this implies for the closed string background we have to understand the conditions implied by the closed string Maurer Cartan equation. It turns out that this equation is difficult to analyze in full generality. Therefore we make an ansatz of the form $\cc=c+\hbar c^{(1)}$, where $c\in A_c$ is a solution of the classical closed string equation of motion and $c^{(1)} \in A_c^{\wedge 2}$. We then find that to lowest order in $\hbar$ the quantum Maurer Cartan equation equation implies that 
\be\no
Q_c[c]\circ f+ f \circ Q_c[c]=1\;\text{,}
\ee
where $Q_c[c]$ is the closed string BRST operator in the classical closed string background $c$ and $f$ is a map,  $f: A_c\to A_c$ construced out of $c^{(1)}$ and $\omega_c$.  In other words, the quantum closed string Maurer Cartan equation implies that $c$ has to be a background where there are no perturbative closed string excitations. This is in agreement with standard argument that open string field theory is inconsistent due to closed string poles arising at the one loop level. Here, this result arises directly form analyzing the Maurer Cartan element for the closed string $IBL_\infty$-algebra.

In the following two sections we define $A_\infty$/ $L_\infty$- and $IBL_\infty$-algebras respectively. Readers  familiar with these algebras may proceed directly to section \ref{hlb} or \ref{sqocLBI} respectively.

\section{$A_\infty$- and $L_\infty$-algebras}\label{sai}
We start by reviewing the construction of $A_\infty$- and $L_\infty$-algebras. Here we establish the notation that will be used throughout the paper. Useful references in the context of $A_\infty$-algebras include 
\cite{Getzler ainfty,Schwarz ainfty} and as a reference for $L_\infty$-algebras we have chosen \cite{Lada linfty}. In the following $A=\bigoplus_{n\in\Z}A_n$ will denote a graded vector space over some field $\mathbb{F}$ (more generally we could consider a module $A$ over some ring $R$). We will use the Koszul sign convention, that is we generate a sign $(-1)^{xy}$ whenever we permute two objects $x$ and $y$. If we permute several object we abbreviate the Koszul sign by $(-1)^\epsilon$.

\subsection{$A_\infty$-algebras}\label{sec:Ainfty}
Following Getzler and Jones, we consider the tensor algebra of $A$
\be\no
TA=\bigoplus_{n=0}^\infty A^{\tp n}\;\text{,}
\ee
and the comultiplication $\Delta:TA\to TA\tp TA$ defined by
\be\no
\Delta (a_1\tp\dots\tp a_n)= \sum_{i=0}^n (a_1\tp \dots\tp a_i)\tp(a_{i+1}\tp\dots\tp a_n)\;\text{.}
\ee
$\Delta$ makes $TA$ a coassociative coalgebra, i.e. 
\be\no
(\Delta\tp\op{id})\circ \Delta = (\op{id}\tp\Delta)\circ \Delta \;\text{.}
\ee
In addition we define the canonical projection maps $\pi_n:TA\to A^{\tp n}$ and inclusion maps $i_n:A^{\tp n}\to TA$.
A coderivation $D\in \op{Coder}(TA)$ is defined by the property 
\be\label{eq:coderTA}
(D\tp \op{id}+\op{id}\tp D)\circ \Delta = \Delta\circ D \;\text{.}
\ee
The defining property $(\ref{eq:coderTA})$ implies that a coderivation $D\in\op{Coder}(TA)$ is uniquely determined by a map $d\in\op{Hom}(TA,A)$, i.e. $\op{Coder}(TA)\cong\op{Hom}(TA,A)$ \cite{Getzler ainfty}. Explicitly the correspondence reads 
\be\no
D\circ i_n= \sum_{i+j+k=n} 1^{\tp i}\tp d_j \tp 1^{\tp k}\;\text{,}
\ee
where $d_n\defineL d\circ i_n$, $1$ denotes the identity map on $A$ and $d=\pi_1\circ D$.
The space of coderivations $\op{Coder}(TA)$ turns out to be a Lie algebra where the Lie bracket is defined by 
\be\no
[D_1,D_2]\defineL D_1\circ D_2 - (-1)^{D_1 D_2}D_2\circ D_1\;\text{.}
\ee
Now an $A_\infty$-algebra is defined by a coderivation $M\in\op{Coder}(TA)$ of degree $1$ (degree $-1$ is considered if $m_1$ is supposed to be a boundary operator rather than a coboundary operator) that squares to zero,
\be\no
M^2=\inv{2}[M,M]=0 \mand |M|=1\;\text{.}
\ee
The corresponding homomorphism is defined by $m=\pi_1\circ M$.
In the case where only $m_1$ and $m_2$ are non-vanishing, we recover the definition of a differential graded associative algebra up to a shift, that is, one takes $sA$ to be the space where the degree is shifted by $1$, i.e. $(sA)_i=A_{i-1}$. The map $s:A\to sA$ has the only effect of changing the degree by $1$. The corresponding inverse map $s^{-1}:sA\to A$ decreases the degree by one. The maps corresponding to the shifted space $sA$ are then defined by
\be\no
\tilde{m}_n\defineL s\circ m_n\circ (s^{-1})^{\tp n}:(sA)^{\tp n} \to sA \;\text{.}
\ee
If only   $\tilde{m}_1$ and $\tilde{m}_2$ are non-vanishing we then recover a differential graded associative algebra on the shifted space. 

Consider now two $A_\infty$-algebras $(A,M)$ and $(A^\prime,M^\prime)$. An $A_\infty$-morphism $F\in\op{Morph}(A,A^\prime)$ from $(A,M)$ to $(A^\prime,M^\prime)$ is defined by 
\be\label{eq:morphTA}
\Delta\circ F = (F\tp F)\circ \Delta \sep F\circ M = M^\prime \circ F \mand |F|=0 \;\text{.}
\ee
The first equation in (\ref{eq:morphTA}) implies that a morphism $F\in\op{Morph}(A,A^\prime)$ is determined by a map $f\in\op{Hom}(TA,A^\prime)$ \cite{Getzler ainfty}. The explicit relation reads
\be\label{eq:morphTAex}
F=\sum_{n=0}^\infty f^{\tp n}\circ \Delta_n \;\text{,}
\ee
where $\Delta_n:TA\to TA^{\tp n}$ denotes the n-fold comultiplication and $f=\pi_1 \circ F$. We use the convention $\Delta_1\defineL\op{id}$. An important property is that the composition of two $A_\infty$-morphisms is again an $A_\infty$-morphism, i.e. for $F\in \op{Morph}(A,A^\prime)$ and $G\in\op{Morph}(A^\prime,A^{\prime\prime})$, $G\circ F\in\op{Morph}(A,A^{\prime\prime})$. This is a direct consequence of equation (\ref{eq:morphTA}).

The concept of Maurer Cartan elements of $A_\infty$-algebras is closely related to that of $A_\infty$-morphisms. 
We define the exponential in $TA$ as 
\be\no
e^a\defineL \sum_{n=0}^\infty a^{\tp n}\;\text{.}
\ee
A Maurer Cartan element $a\in A$ of an $A_\infty$-algebra $(A,M)$ is a degree zero element that satisfies 
\be\no
M(e^a)=0 \qquad \Leftrightarrow \qquad \sum_{n=0}^\infty m_n(a^{\tp n})=0 \;\text{.}
\ee
Note that $\Delta(e^a)=e^a\tp e^a$. Thus we can interpret the exponential $e^a$ of a Maurer Cartan element $a\in A$ as a constant morphism $F\in \op{Morph}(A,A)$, that is $f_0=a$ and $f_n=0$ for all $n\ge1$. Again we used the notation $f_n=f\circ i_n$ and $f\in \op{Hom}(TA,A)$ denotes the homomorphism that corresponds to $F$ (see equation (\ref{eq:morphTAex})). Since we know that the composition of two $A_\infty$-morphisms is again an $A_\infty$-morphism and that a Maurer Cartan element can be interpreted as a constant $A_\infty$-morphism, it follows that an $A_\infty$-morphism maps Maurer Cartan elements into Maurer Cartan elements. The same statement is true for $L_\infty$-algebras (see section \ref{sec:Linfty}).

The language of coderivations is also very useful to describe deformations of $A_\infty$ algebras. Deformations of an $A_\infty$-algebra $(A,M)$ are controlled by the differential graded Lie algebra $\op{Coder}(TA)$ with differential $d_h\defineL [M,\cdot]$ and bracket $[\cdot,\cdot]$. Since $\op{Coder}(TA)\cong \op{Hom}(TA,A)$, $d_h$ and $[\cdot,\cdot]$ have their counterparts defined on $\op{Hom}(TA,A)$, the Hochschild differential and the Gerstenhaber bracket. An infinitesimal deformation of an $A_\infty$-algebra is characterized by the Hochschild cohomology $H^1(d_h,\op{Coder}(TA))$, i.e. the cohomology of $d_h$ at degree $1$. A finite deformation of an $A_\infty$ algebra is an element $D\in \op{Coder}(TA)$ of degree $1$ that satisfies the Maurer-Cartan equation
\be\no
d_h(D)+\inv{2}[D,D]=0 \qquad \Leftrightarrow \qquad (M+D)^2=0\;\text{.}
\ee

We will need one more concept in the context of $A_\infty$-algebras which is called cyclicity. Assume that $A$ is an $A_\infty$-algebra that is additionally endowed with an odd symplectic structure $\omega:A\tp A \to \F$ of degree $-1$. 
We call $d\in\op{Hom}(TA,A)$ cyclic, if the function 
\be\no
\omega(\,d\,,\,\cdot\,) : TA \to \mathbb{F}
\ee
is cyclically symmetric, i.e. 
\be\no
\omega(d_n(a_1,\dots,a_n),a_{n+1})=(-1)^\epsilon \omega(d_n(a_2,\dots,a_{n+1}),a_{1})\;\text{.}
\ee
Since we have the notion of cyclicity for $\op{Hom}(TA,A)$, we also have the notion of cyclicity for $\op{Coder}(TA)$ due to the isomorphism $\op{Coder}(TA)\cong\op{Hom}(TA,A)$. We denote the space of cyclic coderivations by $\op{Coder}^{cycl}(TA)$. An $A_\infty$-algebra $(A,M)$ is called a cyclic $A_\infty$-algebra if $M\in\op{Coder}^{cycl}(TA)$.  It is straightforward to prove that $\op{Coder}^{cycl}(TA)$ is closed w.r.t. the Lie bracket $[\cdot,\cdot]$, and thus we can consider deformations of cyclic $A_\infty$-algebras which are controlled by the differential graded Lie algebra $\op{Coder}^{cycl}(TA)$. The cohomology $H(d_h,\op{Coder}^{cycl}(TA))$ is called cyclic cohomology.

\subsection{$L_\infty$-algebras}\label{sec:Linfty}
Many of the construction in the context of $L_\infty$-algebras are analogous to that of $A_\infty$-algebras. The main difference is that the formulation of $L_\infty$-algebras is based on the graded symmetric algebra $SA$ instead of the tensor algebra $TA$.
The graded symmetric algebra $SA$ is defined as the quotient $TA/I$, where $I$ denotes the two sided ideal generated by the elements $c_1\tp c_2 -(-1)^{c_1c_2}c_2\tp c_1$ with $c_1,c_2\in A$. 
The product $\tp$ defined in $TA$ induces the graded symmetric product $\wedge$ in $SA$.
 The symmetric algebra is the direct sum of the symmetric powers in $A$
\be\no
SA=\bigoplus_{n=0}^\infty A^{\wedge n} \;\text{.}
\ee
All that is simply saying that an element $c_1\wedge \cdots \wedge c_n\in A^{\wedge n}$ is graded symmetric, that is $c_{\sigma_1}\wedge \cdots \wedge c_{\sigma_n}=(-1)^\epsilon c_1\wedge \cdots \wedge c_n$ for any permutation $\sigma \in S_n$ ($S_n$ denotes the permutation group of $n$ elements).

The comultiplication $\Delta:SA\to SA\tp SA$ is defined by
\be\no
\Delta (c_1,\cdots,c_n)=\sum_{i=0}^n\sideset{}{^\prime}\sum_\sigma (c_{\sigma_1}\wedge\cdots\wedge c_{\sigma_i})\tp(c_{\sigma_{i+1}}\wedge\cdots\wedge c_{\sigma_n})\;\text{,}
\ee
where $\sum_\sigma^\prime$ indicates the sum over all permutations $\sigma\in S_n$ constraint to $\sigma_1<\cdots<\sigma_i$ and $\sigma_{i+1}<\cdots<\sigma_n$, i.e. the sum over all inequivalent permutations.

A coderivations $D\in \op{Coder}(SA)$ is defined by
\be\label{eq:coderSA}
(D\tp \op{id}+\op{id}\tp D)\circ \Delta =\Delta \circ D \;\text{.}
\ee
Again the isomorphism $\op{Coder}(SA)\cong \op{Hom}(SA,A)$ holds, and the explicit correspondence between a coderivation $D\in\op{Coder}(SA)$ and its associated map $d=\pi_1\circ D\in\op{Hom}(SA,A)$ is given by 
\cite{Lada linfty}
\be\label{eq:coderSAex}
D\circ i_n= \sum_{i+j=n}\sideset{}{^\prime}\sum_\sigma (d_i\wedge 1^{\wedge j})\circ \sigma\;\text{,}
\ee
where on the right hand side of equation (\ref{eq:coderSAex}) $\sigma$ denotes the map that maps $c_1\wedge \cdots \wedge c_n$ into $(-1)^\epsilon c_{\sigma_1}\wedge \cdots \wedge c_{\sigma_n}$ (again $d_n=d\circ i_n$ and $1$ is the identity map on $A$).

A $L_\infty$-algebra is defined by a coderivation $L\in \op{Coder}(SA)$ of degree $1$ that squares to zero,
\be\no
L^2=0 \mand |L|=1 \;\text{.}
\ee
A $L_\infty$-morphism $F\in \op{Morph}(A,A^\prime)$ from a $L_\infty$-algebra $(A,L)$ to another $L_\infty$-algebra $(A^\prime,L^\prime)$ is defined by
\be\label{eq:morphSA}
\Delta\circ F=(F\tp F)\circ \Delta \sep F\circ L = L^\prime \circ F \mand |F|=0\;\text{.}
\ee
Furthermore it is determined by the map $f=\pi_1\circ F\in \op{Hom}(SA,A^\prime)$ through \cite{Lada linfty}
\be\label{eq:morphSAex}
F=\sum_{n=0}^\infty\inv{n!}f^{\wedge n}\circ \Delta_n \;\text{,}
\ee
where $\Delta_n:SA\to SA^{\tp n}$ denotes the n-fold comultiplication.

Analogous to $A_\infty$-algebras a Maurer Cartan element $c\in A$ of a $L_\infty$-algebra $(A,L)$ is essentially a constant morphism, that is 
\be\no
L(e^c)=0 \mand |c|=0 \;\text{,}
\ee
where the exponential is defined by
\be\no
e^c=\sum_{n=0}^\infty\inv{n!}c^{\wedge n}
\ee
and satisfies $\Delta (e^c) = e^c\tp e^c$.

Finally there is also the notion of cyclicity in the context of $L_\infty$-algebras. Let $(A,L)$ be a $L_\infty$-algebra equipped with an odd symplectic structure $\omega$ of degree $-1$.
We call a coderivation $D\in \op{Coder}(SA)$ cyclic if the corresponding function
\be\no
\omega(\,d\,,\,\cdot\,)
\ee
($d\in\op{Hom}(SA,A)$ is the map corresponding to $D$, see equation (\ref{eq:coderSAex})) is graded symmetric, i.e.
\be\no
\omega(d_n(c_{\sigma_1},\dots,c_{\sigma_n}),c_{\sigma_{n+1}})=(-1)^\epsilon \omega(d_n(c_{1},\dots,c_{n}),c_{n+1}) \;\text{.}
\ee
We denote the space of cyclic coderivations by $\op{Coder}^{cycl}(SA)$.

As a simple illustration of $L_\infty$-morphisms we consider a background shift in closed string field theory. Consider the classical action of closed string field theory, the theory with genus zero vertices $l_{cl}$ only. The corresponding coderivation $L_{cl}$ defines a $L_\infty$-algebra and the action reads (after absorbing $\hbar^{1/2}$ in the string field $c$)
\be
S_{c,cl}=(\omega_c\circ l_{cl})(e^c)\;\text{.}
\ee
Shifting the background simply means that we expand the string field $c$ around $c^\prime$ rather than zero. The action in the new background is $(\omega_c\circ  l_{cl})(e^{c^\prime+c})$. Hence the vertices $ l_{cl}[c^\prime]$ in the shifted background read
\be\no
 l_{cl}[c^\prime]= l_{cl}\circ E(c^\prime)\;\text{,}
\ee
where $E(c^\prime)$ is the map defined by
\be\label{bgs}
E(c^\prime)(c_1\w\dots\w c_n)=e^{c^\prime}\w c_1\w\dots\w c_n\;\text{.}
\ee
In the language of homotopy algebras this shift is implemented by 
\be\label{bgs2}
L_{cl}[c^\prime]=E(-c^\prime)\circ L_{cl}\circ E(c^\prime)\;\text{.}
\ee
Obviously $E(-c^\prime)$ is the inverse map of $E(c^\prime)$. Furthermore $\Delta\circ E(c^\prime)=E(c^\prime)\tp E(c^\prime)$ and therefore $L_{cl}[c^\prime]$ defines also a $L_\infty$-algebra. Thus $E(c^\prime)$ is an $L_\infty$-morphism. In fact, there is a subtlety if the new background does not satisfy the field equations. The initial $L_\infty$-algebra is determined by the vertices $(l_{cl})_n$ where there is no vertex for $n=0$, i.e. $(l_{cl})_0=0$ (A non-vanishing $(l_{cl})_0$ would correspond to a term in the action that depends linearly on the field.). Such an algebra is called a strong $L_\infty$-algebra \cite{Lada linfty}. In the new background we get
\be\no
(l_{cl}[c^\prime])_0=\sum_{n=0}^\infty \inv{n!}(l_{cl})_n({c^\prime}^{\w n})\;\text{,}
\ee
and thus the $L_\infty$-algebra $L_{cl}[c^\prime]$ defines a strong $L_\infty$-algebra only if $c^\prime$ satisfies the field equations \cite{Zwiebach closed}. If this is not the case the resulting algebra is called a weak $L_\infty$-algebra. An odd property of a weak $L_\infty$-algebra $(A,L)$ is that $l_1$  is no longer a differential of $A$, the new relation reads $l_1\circ l_1+l_2\circ(l_0\w 1)=0$.

\section{Homotopy involutive Lie bialgebras}\label{hlb}
The homotopy algebras introduced in the preceding section are suitable for describing the algebraic structures of classical open-closed string field theory as defined in the introduction \cite{Kajiura open-closed}. If one tries to describe quantum open-closed string field theory - the set of vertices satisfying the full quantum BV master equation - in the framework of homotopy algebras, the appropriate language is that of homotopy involutive Lie bialgebras ($IBL_\infty$-algebra). An $IBL_\infty$-algebra is a generalization of a $L_\infty$-algebra.
It is formulated in terms of higher order coderivations - a concept that will be introduced in the next subsection - and requires an auxiliary parameter $x\in \mathbb{F}$ (later on we will identify that parameter with $\hbar$). We will also introduce the notion of morphisms and Maurer Cartan elements in the context of $IBL_\infty$-algebras. Our exposition is based on the work of \cite{Cieliebak ibl}.
In the following we collect their results (in a slightly different notation) to make the paper self-contained.

\subsection{Higher order coderivations}\label{sec:highercoder}
We already know what a coderivation (of order one) on $SA$ is (see equation (\ref{eq:coderSA})). We defined it by an algebraic equation involving the comultiplication $\Delta$. 
The essence of that equation was that a coderivation 
$D\in\op{Coder}(SA)$ is uniquely determined by a homomorphism $d\in\op{Hom}(SA,A)$. Explicitly we had 
\be\label{eq:coderSAex2}
D\circ i_n= \sum_{i+j=n}\sideset{}{^\prime}\sum_\sigma (d_i\wedge 1^{\wedge j})\circ \sigma\;\text{,}
\ee 
where $\pi_1\circ D =d$.

There are two ways to define higher order coderivations.  One is based on algebraic relations like that in equation (\ref{eq:coderSA}) \cite{Markl closed,Akman coder,Bering coder}. 
A coderivation of order two is for example characterized by
\be\no
\Delta_3\circ D -\sideset{}{^\prime}\sum_{\sigma}\sigma\circ (\Delta\circ D\tp \op{id})\circ \Delta + \sideset{}{^\prime}\sum_{\sigma}\sigma\circ (D\tp\op{id}^{\tp 2})\circ \Delta_3=0 \;\text{,}
\ee
where $\sum^\prime_\sigma$ denotes the sum over inequivalent permutations in $S_3$ (the permutation group of three elements) and $\sigma:SA^{\tp 3}\to SA^{\tp 3}$ is the map that permutes the three outputs. For completeness we state the algebraic definition of a coderivation $D\in\op{Coder}^n(SA)$ of order $n$ \cite{Markl closed}
\be\label{eq:highercoder}
\sum_{i=0}^{n}\sideset{}{^\prime}\sum_{\sigma}(-1)^i\sigma\circ(\Delta_{n+1-i}\circ D\tp\op{id}^{\tp i})\circ \Delta_{i+1}=0\;\text{.}
\ee
But similar to the case of a coderivation of order one, this algebraic relation is simply saying that - and this is the alternative definition of higher order coderivations - a coderivation $D\in\op{Coder}^n(SA)$ of order $n$ is uniquely determined by a map $d\in\op{Hom}(SA,\Sigma^nA)$, where $\Sigma^n A=\oplus_{i=0}^n A^{\w i}$. Thus in contrast to a coderivation of order one a coderivation of order $n$ is determined by a linear map on $SA$ with $n$ (and less) outputs rather than just one output. The explicit relation between $D\in\op{Coder}^n(SA)$ and $d\in\op{Hom}(SA,\Sigma^nA)$ is
\be\label{eq:highercoderex}
D\circ i_n=\sum_{i+j=n}\sideset{}{^\prime}\sum_{\sigma}(d_i\w 1^{\w j})\circ \sigma\;\text{,}
\ee
which is the naive generalization of equation (\ref{eq:coderSAex2}). 

A trivial observation is that a coderivation of order $n-1$ is also a coderivation of order $n$ (by simply defining the map with $n$ outputs to be zero), that is
\be\no
\op{Coder}^{n-1}(SA)\subset\op{Coder}^{n}(SA)\;\text{.}
\ee
We call a coderivation $D\in\op{Coder}^n(SA)$ of order $n$ a strict coderivation of order $n$ if the corresponding map $d$ is in $\op{Hom}(SA,A^{\w n})$, that is if the map $d$ has exactly $n$ outputs.
In that case we can identify $d=\pi_n\circ D$.

To continue we define the graded commutator
\be\no
[D_1,D_2]=D_1\circ D_2 -(-1)^{D_1D_2}D_2\circ D_1\;\text{,}
\ee
where $D_1,D_2$ are arbitrary higher order coderivations. Using the defining equations (\ref{eq:highercoder}) it can be shown that \cite{Markl closed} 
\be\label{eq:commhighercoder}
[\op{Coder}^i(SA),\op{Coder}^j(SA)]\subset\op{Coder}^{i+j-1}(SA)\;\text{.}
\ee
In the case $i=j=1$ we recover that $[\cdot,\cdot]$ defines a Lie algebra on $\op{Coder}^1(SA)$, but we see that $[\cdot,\cdot]$ does not define a Lie algebra at higher orders $n>1$.
Of course we can make the collection of all higher order coderivations a Lie algebra, but in the next subsection we will see that there is still a finer structure.

\subsection{$IBL_\infty$-algebra}\label{sec:IBLinfty}
Now we have the mathematical tools to define what an $IBL_\infty$-algebra is. We will furthermore see that one recovers an involutive Lie bialgebra ($IBL$-algebra) as a special case of an $IBL_\infty$-algebra.
$IBL_\infty$-algebras were introduced in \cite{Cieliebak ibl} as well as the notion of $IBL_\infty$-morphisms and Maurer Cartan elements.

Consider the space 
\be\no
\mathfrak{coder}(SA,x)\defineL\bigoplus_{n=1}^\infty x^{n-1}\op{Coder}^{n}(SA)\;\text{,}
\ee
where $x\in \mathbb{F}$ is some auxiliary parameter. An element $\DD\in\coder(SA,x)$ can be expanded
\be\no
\DD= \sum_{n=1}^\infty x^{n-1}D^{(n)}\;\text{,}
\ee
where $D^{(n)}\in\op{Coder}^n(SA)$. In section \ref{summary} we noted that generically, the quantum BV-equation implies that for a given power of $x$, strict coderivations  of order $n> 1$ as well as first order coderivations are present. The latter correspond to Riemann surface of higher genus. 
 In order to take this into account we will now indicate coderivations of order $n$ by a superscript $(n)$ and strict coderivations of order $n$ by a superscript $n$.
We can decompose every coderivation of order $n$ into strict coderivations of order smaller than or equal to $n$. Accordingly, we define the strict coderivation of order $n-g$ corresponding to a coderivation $D^{(n)}$ of order $n$ by
$D^{n-g,g}$, $g\in\{0,\dots,n-1\}$ (later one $g$ will denote the genus). Thus we have
\be\no
D^{(n)}=\sum_{g=0}^{n-1} D^{n-g,g}\;\text{,}
\ee
and $\DD$ expressed in terms of strict coderivations reads
\be\no
\DD=\sum_{n=1}^\infty\sum_{g=0}^\infty x^{n+g-1}D^{n,g}\;\text{.}
\ee
Due to equation (\ref{eq:commhighercoder}) we have
\be\no
[\DD_1,\DD_2]\in \coder(SA,x)\;\text{,}
\ee 
that is, the commutator $[\cdot,\cdot]$ turns $\coder(SA,x)$ into a graded Lie algebra. The space $\coder(SA,x)$ is the Lie algebra on which the construction of $IBL_\infty$-algebras is based.
From a conceptual point of view nothing new happens in the construction of $IBL_\infty$-algebras compared to the construction of $L_\infty$- and $A_\infty$-algebras. The difference is essentially that the underlying objects are more complicated. An $IBL_\infty$-algebra is defined by an element $\LL\in\coder(SA,x)$ of degree $1$ that squares to zero \cite{Cieliebak ibl}:
\be\no
\LL^2=0 \mand |\LL|=1 \;\text{.}
\ee

For completeness we will now  describe $IBL$-algebras as a special case of  $IBL_\infty$-algebras.  Consider an element $\LL\in\coder(SA,x)$ that consists of a strict coderivation of order one and a strict coderivation of order two only:
\be\no
\LL=L^{1,0}+x L^{2,0}\;\text{.}
\ee
Furthermore we restrict to the case where the only non-vanishing components of $l^{1,0}\defineL \pi_1\circ L^{1,0}:SA\to A$ and $l^{2,0}\defineL \pi_2\circ L^{2,0}:SA\to A^{\w 2}$ are
\be\no
\tilde{d}\defineL l^{1,0}\circ i_1 :A\to A  \sep \tilde{[\cdot,\cdot]}\defineL l^{1,0}\circ i_2 :A^{\w 2}\to A\;\text{,}
\ee
\be\no
\qquad\tilde{\delta}\defineL l^{2,0}\circ i_1:A\to A^{\w 2}\;\text{.}
\ee
To recover the definition of an involutive Lie bialgebra we have to shift the degree by one (see section \ref{sai}), i.e. we define maps on the shifted space $sA$ by
\be\no
d\defineL s\circ \tilde{d}\circ s^{-1} \sep [\cdot,\cdot]\defineL s\circ \tilde{[\cdot,\cdot]}\circ {(s^{-1})}^{\w 2}\;\text{,}
\ee
\be\no
\delta\defineL s^{\w 2}\circ \tilde{\delta}\circ s^{-1}\;\text{.}
\ee
The requirement $\LL^2=0$ is then equivalent to the seven conditions 
\begin{align}\label{eq:IBL}
\phantom{\sum_{\sigma}}\tilde{d}^2=0 &\quad\Leftrightarrow\quad \text{$d$ is a differential}\\\no
\phantom{\sum_{\sigma}}\tilde{d}\circ\tilde{[\cdot,\cdot]}+\tilde{[\cdot,\cdot]}\circ (\tilde{d}\w 1 +1 \w \tilde{d}) &\quad\Leftrightarrow\quad \text{$d$ is a derivation over $[\cdot,\cdot]$} \\\no
\phantom{\sum_{\sigma}}(\tilde{d}\w 1+1\w \tilde{d})\circ\tilde{\delta}+ \tilde{\delta}\circ\tilde{d}  &\quad\Leftrightarrow\quad \text{$d$ is a coderivation over $\delta$}\\\no
\sideset{}{^\prime}\sum_{\sigma}\tilde{[\cdot,\cdot]} \circ (\tilde{[\cdot,\cdot]}\w 1)\circ \sigma =0  &\quad\Leftrightarrow\quad \text{$[\cdot,\cdot]$ satisfies the Jacobi identity}\\\no
\phantom{\sum_{\sigma}}(\tilde{\delta}\w1+1\w\tilde{\delta})\circ\tilde{\delta}=0 &\quad\Leftrightarrow\quad \text{$\delta$ satisfies the co-Jacobi identity}\\\no
\sideset{}{^\prime}\sum_{\sigma}(\tilde{[\cdot,\cdot]}\w 1)\circ \sigma \circ (\tilde{\delta}\w1+1\w\tilde{\delta})+ \tilde{\delta}\circ\tilde{[\cdot,\cdot]}=0   
&\quad\Leftrightarrow\quad \text{compatibility of $\delta$ and $[\cdot,\cdot]$}\\\no
\phantom{\sum_{\sigma}}\tilde{[\cdot,\cdot]}\circ\tilde{\delta}=0  &\quad\Leftrightarrow\quad \text{involutivity of $\delta$ and $[\cdot,\cdot]$}\;\text{.}
\end{align}
These are just the conditions defining a differential involutive Lie bialgebra.  

\subsection{$IBL_\infty$-morphisms and Maurer Cartan elements}\label{sec:IBLinftymorph}
A $L_\infty$-morphism was defined by two equations (\ref{eq:morphSA}). The first involves the comultiplication and implies that a $L_\infty$-morphism can be expressed by a homomorphism from $SA$ to $A$ 
(\ref{eq:morphSAex}), i.e. it determines its structure. We do not of know a suitable generalization of that equation to the case of $IBL_\infty$-algebras, but instead one can easily generalize equation (\ref{eq:morphSAex}). 
The second equation is just saying that the morphism commutes with the differentials and looks identically in the case of $IBL_\infty$-algebras.\\
Let $(A,\LL)$ and $(A^\prime,\LL^\prime)$ be two $IBL_\infty$-algebras. 
An $IBL_\infty$-morphism $\FF\in\morph(A,A^\prime)$ is defined by \cite{Cieliebak ibl}
\be\label{eq:morphIBLex}
\FF=\sum_{n=0}^\infty \inv{n!}\mathcal{F}^{\w n}\circ \Delta_n \sep \FF\circ \LL=\LL^\prime\circ\FF \mand |\FF|=0 \;\text{,}
\ee
where 
\be\no
\mathcal{F}=\sum_{n=0}^\infty x^{n-1}F^{(n)} \mand F^{(n)}:SA\to \Sigma^nA^\prime\;\text{.}
\ee
Recall that $\Sigma^nA^\prime=\oplus_{i=1}^n{A^\prime}^{\w i}$. Thus we can decompose $F^{(n)}$ into a set of maps $F^{n-g,g}:SA\to {A^\prime}^{\w n-g}$, $g\in\{0,\dots,n-1\}$ (in the same way we decomposed higher order coderivations).
Expressed in terms of $F^{n,g}$ we have
\be\label{eq:morphIBLmap}
\mathcal{F}=\sum_{n=1}^\infty\sum_{g=0}^\infty x^{n+g-1}F^{n,g}\;\text{.}
\ee
Due to the lack of an algebraic relation governing the structure of an $IBL_\infty$-morphism - an equation generalizing (\ref{eq:morphSA}) - it is not obvious that the composition of two morphisms yields again a morphism. Nevertheless, in \cite{Cieliebak ibl} this has been shown to be true. 

To complete the section we finally state what a Maurer Cartan element of an $IBL_\infty$-algebra $(A,\LL)$ is.
Let $c^{n,g}\in A^{\w n}$ be of degree zero. $\cc=\sum_{n=1}^\infty\sum_{g=0}^\infty x^{n+g-1}c^{n,g}$ is called a Maurer Cartan element of $(A,\LL)$ if \cite{Cieliebak ibl}
\be\no
\LL(e^\cc)=0\;\text{,}
\ee
that is we can interpret a Maurer Cartan element as a constant morphism on $(A,\LL)$ (Here the exponential is the same as in the case of $L_\infty$-algebras, i.e. $e^\cc=\sum_{n=0}^\infty \inv{n!}\cc^{\w n}$.).

\section{Quantum open-closed homotopy algebra}\label{sqocLBI}
After all the preliminary parts about homotopy algebras, we now turn to string field theory and show how these mathematical structures are realized therein. 
At the classical level the spaces of open- and closed strings, $A_o$ and $A_c$ are vector spaces over the field $\C$ but at the quantum level $A_o$ and $A_c$ become a module over the Grassmann numbers $\C_{\Z_2}=\C_0\oplus\C_1$, where $\C_0$ resp. $\C_1$ represents the commuting resp. anticommuting numbers. That is at the quantum level we have to allow for both bosonic and fermionic component fields of the string field and the space of string fields becomes a bigraded space. The ghost number grading is denoted by $|\cdot|_{gh}$ whereas the Grassmann grading is denoted by $|\cdot|_{gr}$. We define a total $\Z_2$ grading by $|\cdot|=|\cdot|_{gh}+|\cdot|_{gr} \; mod \; 2$. The string fields $c$ and $a$ are of total degree zero, i.e. we pair ghost number even with Grassmann even and ghost number odd with Grassmann odd.

It turns out to be convenient to express the vertices with open and closed inputs (\ref{ocv}) in terms of homomorphisms with outputs in the shifted open string space $s^{-1}A_o$ introduced in section \ref{sai}, that is 
\be\no
\sum_{n}\sum_{m_1,\dots,m_b}\hbar^{2g+b+n/2-1}\;\mathcal{V}^{b,g}_{n,m_1,\dots,m_b}(c,a)
= \inv{b!}\hbar^{2g+b-1} (\tilde{\omega}_o^{\tp b}\circ \tilde{f}^{b,g})(e^{\hbar^{1/2}c};\underbrace{\bar{e}^a,\dots,\bar{e}^a}_{\text{$b$ times}})\;\text{,}
\ee
where $\tilde{f}^{b,g}\in \op{Hom}(SA_c,\C_{\Z_2})\tp (\op{Hom}^{cycl}(TA_o,s^{-1}A_o))^{\w b}$ and $\tilde{\omega}_o=\omega_o\circ s$. Note that  $\bar{e}^a\defineL\sum_{n=1}^\infty \inv{n}a^{\tp n}$ deviates from the definition of $e^a=\sum_{n=0}^\infty a^{\tp n}$ in section \ref{sai}, but the symmetry factor of $1/n$ turns out to be convenient later on.  There are actually two reasons why we defined this expression in terms of $\tilde{\omega}_o$ rather than $\omega_o$. The first is purely technical, namely that  $\tilde{\omega}_o$ and hence $\tilde{f}^{b,g}$ are of degree zero so that we do not have to worry about signs whenever we permute these objects. The second is that we finally want to interpret the collection of all maps $\tilde{f}^{b,g}$ as the defining map of an $IBL_\infty$-morphism (see equation (\ref{eq:morphIBLmap})) which is by definition of degree zero. To summarize, the full BV quantum action of open-closed string field theory can be expressed as
\be\label{eq:BVaction}
S=\sum_{g=0}^\infty \hbar^{2g-1}(\omega_c\circ l^g)(e^{\hbar^{1/2}c})
+ \sum_{b=1}^\infty\sum_{g=0}^\infty \hbar^{2g+b-1}(\tilde{\omega}^{\tp b}_o\circ \tilde{f}^{b,g})(e^{\hbar^{1/2}c};\underbrace{\bar{e}^a,\dots,\bar{e}^a}_{\text{$b$ times}})\;\text{.}
\ee
The idea that the set of all vertices with open and closed inputs can be interpreted as a morphism between appropriate homotopy algebras came up in \cite{Kajiura open-closed,Kajiura open-closed2} for the classical open-closed string field theory, where one considers only vertices with genus zero and at most one boundary:\\
It is known that the vertices of classical closed string field theory define a $L_\infty$-algebra $(A_c,L_{cl})$ \cite{Zwiebach closed}. On the other hand, the vertices of classical open string field theory define an $A_\infty$-algebra $(A_o,M_{cl})$ \cite{Zwiebach open1, Zwiebach open2, Kajiura open}, which makes the space $\op{Coder}^{cycl}(TA_o)$ - the cyclic Hochschild complex - a differential graded Lie algebra (see section \ref{sec:Ainfty}). Since a DGL is a special case of a $L_\infty$-algebra up to a shift in degree - the $L_\infty$-algebra as defined here, is realized on $s^{-1}\op{Coder}^{cycl}(TA_o)$-  the set of open and closed vertices can be identified as a $L_\infty$-morphism between the $L_\infty$-algebra of closed strings and the DGL on the cyclic Hochschild complex of open strings
\be\label{eq:OCHAform}
(A_c,L_{cl})\xrightarrow[]{L_\infty-\text{morphism}} (\op{Coder}^{cycl}(TA_o),d_h,[\cdot,\cdot])\;\text{.}
\ee
This is the open-closed homotopy algebra of Kajiura and Stasheff \cite{Kajiura open-closed,Kajiura open-closed2}.

In order to generalize this picture to the quantum level, we first have to identify the new structures on the closed string and on the open string side of (\ref{eq:OCHAform}), i.e. the algebraic structure on $A_c$ and 
$\op{Coder}^{cycl}(TA_o)$. This is the content of the following two subsections. In the last part of this section we will connect the open and closed string part by an $IBL_\infty$-morphism and finally define the quantum open-closed homotopy algebra - the algebraic structure of quantum open-closed string field theory. 

\subsection{Loop homotopy algebra of closed strings}\label{sec:loop}
The reformulation of the algebraic structures of closed string field theory in terms of homotopy algebras has been done in \cite{Markl closed} and  will briefly review the results here. The corresponding homotopy algebra is called loop algebra.

The space of closed string fields $A_c$ is endowed with an odd symplectic structure $\omega_c$. Choose a homogeneous basis $\{e_i\}$ of $A_c$, where we denote the degree of $e_i$ by $|e_i|=i$. We use DeWitt's sign convention \cite{Witt super}, that is, we introduce for every basis vector $e_i$ the vector ${}_ie\defineL (-1)^ie_i$. Einstein's sum convention is modified in that we sum over repeated indices whenever one of the indices is an upper left resp. right index and the other one is a lower right resp. left index. A vector $c\in A_c$ can be expanded in terms of the left or the right basis, i.e. 
\be
c=c^i\;{}_ie=e_i\;{}^ic \;\text{,}
\ee
and the expansion coefficients are related via ${}^ic=(-1)^{|c|i}\,c^i$.
Let $\{e^i\}$ be its dual basis with respect to the symplectic structure $\omega_c$, i.e. 
\be
\omega_c({}_ie,e^j)={}_i\delta^j \;\text{,}
\ee
where ${}_i\delta^j $ denotes the Kronecker delta. Note that $e^i$ has degree $1-i$ and hence ${}^ie=(-1)^{i+1}e^i$. These definitions ensure that $\omega_c({}^je,e_i)=\omega_c({}_ie,e^j)={}_i\delta^j$. 
$\omega_c$ regarded as a map from $A_c$ to $A_c^\ast$ is invertible and we denote its inverse by $\omega_c^{-1}$. It follows that $\omega_c^{-1}=\inv{2}e_i\w e^i \in A^{\w 2}$ and $|\omega_c^{-1}|=1$. We can lift $\omega_c^{-1}$ to a strict coderivation $D(\omega_c^{-1})\in \op{Coder}^2(SA_c)$ of order two defined by
\be\label{eq:coderomega}
\pi_1\circ D(\omega_c^{-1})=0 \mand \pi_2 \circ D(\omega_c^{-1})=\omega_c^{-1}\;\text{,}
\ee
Utilizing the isomorphism $\op{Hom}^{cycl}(SA_c,A_c)\cong\op{Coder}^{cycl}(SA_c)$ we can lift  the closed string vertices $l^g\in \op{Hom}^{cycl}(SA_c,A_c)$ of the BV action (\ref{eq:BVaction}) to a coderivation $L^g\in\op{Coder}^{cycl}(SA_c)$, $g\in\N_0$. The combination
\be\label{eq:closedIBL}
\LL_c= \sum_{g=0}^\infty \hbar^g L^g +  \hbar  D(\omega_c^{-1})
\ee
defines an element in $\coder(SA_c,\hbar)$ of degree $1$. The homotopy loop algebra of closed string field theory is defined by \cite{Markl closed}
\be\label{eq:loop}
\LL_c^2=0\;\text{.}
\ee 
Thus the loop algebra is a special case of an $IBL_\infty$-algebra. Furthermore equation (\ref{eq:loop}) is equivalent to the following statements:
\be\label{eq:main}
\sum_{g_1+g_2=g \atop i_1+i_2=n}\sideset{}{^\prime}\sum_\sigma l^{g_1}_{i_1+1}\circ (l^{g_2}_{i_2}\w1^{\w i_1})\circ \sigma + l^{g-1}_{n+2}\circ (\omega_c^{-1}\w 1^{\w n})=0
\ee
\be\label{eq:cyclic}
e_i\w l^{g}_{n+1}\circ({}^ie\w1^{\w n})=0
\ee
Equation (\ref{eq:cyclic}) is merely saying that $l^g$ has to be cyclic whereas equation (\ref{eq:main}) is called the main identity \cite{Zwiebach closed}.
These are the algebraic relations of quantum closed string field theory expressed in terms of homotopy algebras, i.e. the algebraic relations of the loop algebra are equivalent to the BV equation with closed strings only. 

\subsection{IBL structure on cyclic Hochschild complex}
In section \ref{sec:Ainfty} we already saw that the space of cyclic coderivations $\op{Coder}^{cycl}(TA)$ is a Lie algebra, with Lie bracket $[D_1,D_2]=D_1\circ D_2-(-1)^{D_1D_2}D_2\circ D_1$. If $A$ is in addition an $A_\infty$-algebra $(A,M)$, the space $\op{Coder}^{cycl}(TA)$ becomes a DGL where the differential is defined by $d_h=[M,\cdot]$. But it turns out that we can make $\op{Coder}^{cycl}(TA)$ even a differential involutive Lie bialgebra, i.e. there is a map $\delta:\op{Coder}^{cycl}(TA)\to \op{Coder}^{cycl}(TA)^{\tp 2}$ such that $d_h$, $[\cdot,\cdot]$ and $\delta$ satisfy the defining equations (\ref{eq:IBL}) of an IBL-algebra.
Recall that $\op{Coder}^{cycl}(TA)\stackrel{\pi_1}\cong \op{Hom}^{cycl}(TA,A)\stackrel{\mathrm{\omega}}\cong \op{Hom}^{cycl}(TA,\F)$, so if we succeed to define such a map $\delta$ on $\op{Hom}^{cycl}(TA,\F)$ there will be a corresponding map on $\op{Coder}^{cycl}(TA)$ satisfying the same properties. Following \cite{Chen liebi,Cieliebak ibl} we define $\delta:\op{Hom}^{cycl}(TA,\F)\to \op{Hom}^{cycl}(TA,\F)^{\tp 2}$ by
\begin{align}\label{eq:delta}
(\delta f)&(a_1,\dots,a_n)(b_1,\dots,b_m)\\\no
\defineL&\sum_{i=0}^{n-1}\sum_{j=0}^{m-1}(-1)^\epsilon f(\dot{e}_p,a_{i+1},\dots,a_n,a_1,\dots,a_i,\dot{e}^p,b_{j+1},\dots,b_m,b_1,\dots,b_j)\;\text{,}
\end{align}
where $\{\dot{e}_p\}$ is a basis of $A$ with its index lifted by $\omega_0$.  This definition ensures that $\delta f$ is cyclic and graded symmetric. Furthermore,  $d_h$, $[\cdot,\cdot]$ and $\delta$ satisfy all conditions of (\ref{eq:IBL}) \cite{Chen liebi,Cieliebak ibl}. We will call the corresponding map defined on $\op{Coder}^{cycl}(TA)$ also $\delta:\op{Coder}^{cycl}(TA)\to\op{Coder}^{cycl}(TA)^{\tp 2}$. Note that this map has degree $2$. Now let us put this into the language of $IBL_\infty$-algebras. First we shift the degree by one
\be\no
\tilde{d_h}\defineL s^{-1}\circ d_h\circ s \sep \tilde{[\cdot,\cdot]}\defineL s^{-1}\circ[\cdot,\cdot]\circ s^{\tp 2}\;\text{,}
\ee
\be\no
\tilde{\delta}=(s^{-1})^{\tp 2}\circ\delta\circ s\;\text{,}
\ee 
and then we lift these maps separately to coderivations on $S\mathcal{\ti{A}}$ where $\mathcal{\ti{A}}=s^{-1}\op{Coder}^{cycl}(SA)$ (see section \ref{sec:IBLinfty}):
\be\no
\hati{d_h} \in \op{Coder}(S\mathcal{\ti{A}})\sep \hati{[\cdot,\cdot]} \in \op{Coder}(S\mathcal{\ti{A}})\sep \hati{\delta} \in \op{Coder}^2(S\mathcal{\ti{A}})\;\text{.}
\ee
The coderivations lifted from a map are indicated by a hat, whereas the tilde refers to the shift in degree.
The statement that the maps $d_h$, $[\cdot,\cdot]$ and $\delta$ satisfy the defining relations of a differential $IBL$-algebra is then equivalent to
\be\no
(\hati{d_h}+\hati{[\cdot,\cdot]}+x\hati{\delta})^2=0\;\text{.}
\ee
If the algebra $A$ is not endowed with the structure of an $A_\infty$-algebra the differential $d_h$ is absent, but still we have an $IBL$-algebra defined by
\be\no
\ti{\LL}_o^2=0\;\text{.}
\ee
where 
\be\label{eq:openIBL}
\ti{\LL}_o\defineL \hati{[\cdot,\cdot]}+x\hati{\delta} \mand  |\ti{\LL}_o|=1 \;\text{.}
\ee
This is the structure that will enter in the definition of the quantum open-closed homotopy algebra. 
That means that we do not anticipate that the vertices of classical open string field theory define an $A_\infty$-algebra but rather derive it from the quantum open-closed homotopy algebra.

\subsection{Quantum open-closed homotopy algebra}\label{sec:QOCHA}
Now we can put the parts together and define the quantum open-closed homotopy algebra (QOCHA). The QOCHA is defined by an $IBL_\infty$-morphisms from the $IBL_\infty$-algebra of closed strings to the $IBL$-algebra of open strings
\be\label{eq:QOCHApic}
(A_c,\LL_c)\xrightarrow[]{IBL_\infty-\text{morphism}} (s^{-1}\op{Coder}^{cycl}(TA_o),\ti{\LL}_o)\;\text{,}
\ee
where $\LL_c\in\coder(SA_c,\hbar)$ is defined in equation (\ref{eq:closedIBL}) and $\ti{\LL}_o\in\coder(S\ti{\mathcal{A}}_o,\hbar)$ is defined in equation (\ref{eq:openIBL}). We use the abbreviation 
$\ti{\mathcal{A}}_o=s^{-1}\op{Coder}^{cycl}(TA_o)$. More precisely we have an $IBL_\infty$-morphism $\ti{\FF}\in \morph(A_c,\ti{\mathcal{A}}_o)$, that is
\be\label{eq:QOCHA}
\ti{\FF}\circ\LL_c=\ti{\LL}_o\circ\ti{\FF} \mand |\ti{\FF}|=0.
\ee
The convention here is that we put a tilde on every map whose domain or target space is $\ti{\mathcal{A}_o}$.
The morphism $\ti{\FF}$ is determined by a map $\ti{\mathcal{F}}$ through (see equation (\ref{eq:morphIBLex}) and (\ref{eq:morphIBLmap}))
\be\no
\ti{\FF}=\sum_{n=0}^\infty \inv{n!}\ti{\mathcal{F}}^{\w n}\circ\Delta_n\;\text{,}
\ee
where
\be\no
\ti{\mathcal{F}}=\sum_{b=1}^\infty\sum_{g=0}^\infty \hbar^{g+b-1}\ti{F}^{b,g}\;\text{,}
\ee
and 
\be\no
\ti{F}^{b,g}:SA_c\to \ti{\mathcal{A}}_o^{\w b}\;\text{.}
\ee
Again we can utilize the isomorphism $\op{Coder}^{cycl}(TA_o)\cong \op{Hom}^{cycl}(TA_o,A_o)$ induced by the projection map $\pi_1:TA_o\to A_o$ to extract the maps determining $\ti{\mathcal{F}}$ and hence $\ti{\FF}$:
\be\no
\ti{f}^{b,g}\defineL\pi^{\w b}_1 \circ \ti{F}^{b,g} \;:\; SA_c\tp TA_o^{\tp b}\to s^{-1}A_o^{\w b}\; \text{.}
\ee
It turns out that (\ref{eq:QOCHA}) together with (\ref{eq:closedIBL})  and (\ref{eq:openIBL}) is equivalent to the algebraic constraints imposed by the BV master equation (\ref{eq:BVeqsummary}) for the vertices in the action of open-closed string field theory provided we  identify the maps $l^g$ and $\ti{f}^{b,g}$ with the closed- and open-closed vertices of the BV action $S$  in (\ref{eq:BVaction}). The detailed proof of this equivalence is postponed to  appendix \ref{app:BV}. Schematically the equivalence goes as follows:  The BV operator $\Delta^{BV}$  is a second order derivation on the space of functions (e.g.  \cite{Schwarz bv, Getzler bv}), whereas the odd Poisson bracket $(\cdot,\cdot)$ and the action $S$ together define a derivation $(S,\cdot)$ on the space of functions. More precisely, the BV operator and the odd Poisson bracket split into open and closed parts: 
\be\no
\Delta^{BV}=\Delta_o^{BV}+\Delta_c^{BV} \mand (\cdot,\cdot)=(\cdot,\cdot)_o+(\cdot,\cdot)_c
\ee
The counterpart of the open string BV operator $\Delta_o^{BV}$ is the second order coderivation $\hati{\delta}$ and the derivation $(S,\cdot)_o$ translates into the coderivation $\hati{[\cdot,\cdot]}$. In fact this is not quite correct since the BV operator $\Delta_o^{BV}$ will also partly play the role of $\hati{[\cdot,\cdot]}$. The reason for this is that $\Delta_o^{BV}$ is not a strict second order derivation in contrast to $\hati{\delta}$. On the closed string side a similar identification holds. The counterpart of the closed string BV operator $\Delta_c^{BV}$ is $D(\omega_c^{-1})$ (see equation (\ref{eq:coderomega})) and that of the derivation $(S,\cdot)_c$ is the coderivation $\sum_g\hbar^{g}L^g$ of equation (\ref{eq:closedIBL}). Again this is just the naive identification since $(S,\cdot)_c$ partly translates into $D(\omega_c^{-1})$.

In order to gain a better geometric intuition of  (\ref{eq:QOCHA}) it is useful to disentangle this equation. 
First consider the left hand side of equation (\ref{eq:QOCHA}). We have
\be\label{pd1}
\Delta_n\circ L^g= \sum_{i+j=n-1}(\op{id}^{\tp i}\tp L^g \tp \op{id}^{\tp j})\circ \Delta_n
\ee
and 
\begin{align}\no\label{pd2}
\Delta_n \circ D(\omega_c^{-1})=& \sum_{i+j=n-1}\bracket{\op{id}^{\tp i}\tp D(\omega_c^{-1}) \tp \op{id}^{\tp j}}\circ \Delta_n\\
  +& \sum_{i+j+k=n-2}\bracket{\op{id}^{\tp i}\tp D(e_i)\tp \op{id}^{\tp j}\tp D(e^i)\tp \op{id}^{\tp k}}\circ \Delta_n \;\text{,}
\end{align}
where $D(e_i)$ denotes the coderivation of order one defined by
\be\no
\pi_1\circ D(e_i)= e_i\;\text{.}
\ee
In the following we abbreviate $L_q=\sum_{g}\hbar^gL^g$. We get 
\begin{align}\no
\ti{\FF}\circ \LL_c =&\phantom{+}\sum_{n=0}^\infty \inv{n!}\sum_{i+j=n-1}(\cF^{\w i}\w \cF\circ(L_q+\hbar D(\omega_c^{-1})) \w \cF^{\w j})\circ \Delta_n\\\no
 &+\sum_{n=0}^\infty \inv{n!}\sum_{i+j+k=n-2}\hbar\bracket{\cF^{\w i}\w\cF\circ D(e_i)\w \cF^{\w j}\w \cF \circ D(e^i)\w \cF^{\w k}}\circ \Delta_n\\\no
 =& \bracket{\bracketii{\cF\circ \LL_c+\inv{2}\hbar (\cF \circ D(e_i)\w \cF \circ D(e^i))\circ\Delta }\w \ti{\FF}}\circ \Delta \;\text{.}
\end{align}
Now let us turn to the right hand side of equation (\ref{eq:QOCHA}). There we have the maps $\hati{\delta}$ and $\hati{[\cdot,\cdot]}$. The defining map $\ti{\delta}=\pi_2\circ \hati{\delta}$ of $\hati{\delta}$ has two outputs and one input. Recall that we defined the order of a coderivation by the number of outputs of the underlying defining map (see section \ref{sec:highercoder}). Similarly we can define higher order derivations by the number of inputs of the underlying defining map \cite{Markl closed}. So we can interpret $\hati{\delta}$ either as a second order coderivation or a first order derivation and $\hati{[\cdot,\cdot]}$ as a first order coderivation or a second order derivation. For our purpose here the second point of view will prove useful. Having these properties in mind one can show that 
\be\no
\hati{\delta}\circ\ti{\FF}=\bracket{\hati{\delta}\circ\cF\w \ti{\FF}}\circ\Delta
\ee
and\footnote{Equation (\ref{eq:BVanalog}) can be derived in analogy to $\Delta^{BV}e^S=(\Delta^{BV}S+\inv{2}(S,S))e^S$, where $(f,g)\defineL (-1)^f(\Delta^{BV}(fg)-(\Delta^{BV}f)g-(-1)^f(\Delta^{BV}g))$,
in the BV formalism (see \cite{Getzler bv} and appendix \ref{app:BV} for more details).}
\be\label{eq:BVanalog}
\hati{[\cdot,\cdot]}\circ \ti{\FF}=\bracketiii{\bracketii{  \hati{[\cdot,\cdot]}\circ\cF +\inv{2}\hati{[\cdot,\cdot]}\circ \bracketi{\cF\w \cF}\circ\Delta - \bracketi{(\hati{[\cdot,\cdot]}\circ\cF) \w \cF}\circ\Delta  }   \w \ti{\FF}}\circ\Delta\;\text{.}
\ee
Besides the properties of $\hati{\delta}$ and $\hati{[\cdot,\cdot]}$, we also used cocommutativity and coassociativity of $\Delta$. 
Thus we can equivalently define the QOCHA by
\begin{align}\label{eq:QOCHAex}
&\cF\circ\LL_c+\frac{\hbar}{2} \bracketi{\cF\circ D(e_i)\w\cF\circ D(e^i)}\circ\Delta \\\no
 &=\ti{\LL}_o\circ \cF + \inv{2}\hati{[\cdot,\cdot]}\circ\bracketi{\cF\w \cF}\circ \Delta -\bracket{(\hati{[\cdot,\cdot]}\circ\cF)\w \cF}\circ \Delta\;\text{.}
 \end{align}
This is the equation we will match with the quantum BV master equation in appendix \ref{app:BV}.
Furthermore, the individual terms in equation (\ref{eq:QOCHAex}) can be identified with the five distinct sewing operations 
of bordered Riemann surfaces with closed string insertions (punctures in the bulk) and open string insertions (punctures on the boundaries) defined in \cite{Zwiebach open-closed}.
The sewing either joins two open string insertions or two closed string insertions. In addition the sewing may involve a single surface or two surfaces.
\bi
\item[(i)] Take an open string insertion of one surface and sew it with another open string insertion on a second surface.
The genus of the resulting surface is the sum of the genera of the indiviual surfaces, whereas the number of boundaries decreases by one.
This operation is identified with
\be\no
\inv{2}\hati{[\cdot,\cdot]}\circ\bracketi{\cF\w \cF}\circ \Delta -\bracket{(\hati{[\cdot,\cdot]}\circ\cF)\w \cF}\circ \Delta\;\text{.}
\ee
\item[(ii)] Sewing of two open string insertions living on the same boundary. This operation obviously increases the number of boundaries by one but leaves the genus unchanged.
It is described by
\be\no
\hati{\delta}\circ\cF \;\text{,}
\ee
in the homotopy language.
\item[(iii)] Consider a surface with more than one boundary. Take an open string insertion of one boundary and sew it with another open string insertion on a second boundary.
This operation increases the genus by one and decreases the number of boundaries by one. It is identified with
\be\no
\hati{[\cdot,\cdot]}\circ \cF \;\text{.}
\ee
\item[(iv)] Sewing of two closed string insertion, both lying on the same surface. This attaches a handle to the surface and hence increases the genus by one, whereas the number of boundaries does not change.
We identify it with
\be\no
 \cF\circ  D(\omega_c^{-1}) \;\text{.}
\ee
\item[(v)] Take a closed string insertion of one surface and sew it with another closed string insertion on a second surface.
The genus and the number of boundaries of the resulting surface is the sum of the genera and the sum of the number of boundaries respectively of the input surfaces.
The sewing where both surfaces have open and closed insertions is identified with
\be\no
\bracketi{\cF\circ D(e_i)\w\cF\circ D(e^i)}\circ\Delta\;\text{,}
\ee
whereas the sewing involving a surface with closed string insertions only and another surface with open and closed string insertions is identified with
\be\no
\cF \circ L_q\;\text{.}
\ee
\ei
This provides the geometric interpretation of all individual terms in (\ref{eq:QOCHAex}). 

Let us now focus on the vertices with open string insertions only. These vertices are also comprised in the $IBL_\infty$-morphism $\ti{\FF}$ and defined by setting the closed string inputs to zero. More precisely, let $\cM=\cF |$  be the restriction of $\cF$ onto the subspace $A_c^{\wedge 0}$ without closed strings. The weighted sum of open string vertices is then given by
\be\no
\cM=\sum_{b=1}^\infty\sum_{g=0}^\infty \hbar^{g+b-1}\ti{M}^{b,g} \sep  \ti{M}^{b,g}\in \ti{\mathcal{A}}_o^{\w b}\;\text{,}
\ee
 where $\ti{M}^{b,g}=\ti{F}^{b,g}|$. The complement of $\cM$ - the vertices with at least one closed string input - is denoted by $\cN$, so that
 \be\label{eq:splitopen}
 \cF=\cM+\cN\;\text{.}
 \ee
 
In the classical limit $\hbar\to0$ we expect to recover the OCHA defined by Kajiura and Stasheff \cite{Kajiura open-closed,Kajiura open-closed2}. Indeed the $IBL_\infty$-morphism $\ti{\FF}$ reduces to a $L_\infty$-morphism, the loop algebra $\LL_c$ of closed strings reduces to a $L_\infty$-algebra $L_{cl}\defineL L^0$ and the $IBL$-algebra on the space of cyclic coderivations becomes an ordinary Lie algebra. The defining equation (\ref{eq:QOCHAex}) of the QOCHA simplifies to
\be\no
\ti{F}_{cl}\circ L_{cl}=\inv{2}\ti{[\cdot,\cdot]}\circ (\ti{F}_{cl}\w\ti{F}_{cl})\circ\Delta \;\text{,}
\ee
and reads
\be\label{eq:OCHA2}
F_{cl}\circ L_{cl}=\inv{2}[F_{cl},F_{cl}]\circ \Delta
\ee
in the unshifted space (i.e. $F_{cl}\in\op{Coder}^{cycl}(TA_o)$ and $|F_{cl}|=1$).
$F_{cl}\defineL F^{1,0}$ is the map of $\mathcal{F}$ with one boundary and genus zero and the corresponding $L_\infty$-morphism is given by $\sum_{n}\inv{n!}\ti{F_{cl}}^{\w n}\circ\Delta_n$ (see section \ref{sec:Linfty}).
Separating the purely open string vertices $M_{cl}$ from $F_{cl}$, we see that those have to satisfy an $A_\infty$-algebra (since $L_{cl}|=0$), i.e. they define a classical open string field theory \cite{Zwiebach open1,Zwiebach open2}. Thus the space $\op{Coder}^{cycl}(TA_o)$ turns into a DGL with differential $d_h=[M_{cl},\cdot]$ and equation (\ref{eq:OCHA2}) finally reads 
\be\label{eq:OCHA}
N_{cl}\circ L_{cl}= d_h(N_{cl})+\inv{2}[N_{cl},N_{cl}]\circ \Delta \;\text{,}
\ee
where $N_{cl}=F_{cl}-M_{cl}$ denotes the vertices with at least one closed string input. Eqn (\ref{eq:OCHA}) is precisely the OCHA defined in \cite{Kajiura open-closed,Kajiura open-closed2}. The physical interpretation of  $N_{cl}$ is that it describes the deformation of open string field theory by turning on a closed string background.  The vanishing of the r.h.s. is the condition for a consistent classical field theory while the l.h.s. vanishes if the closed string background solves the classical closed string field theory equation of motion. Eqn (\ref{eq:OCHA}) then implies that the open-closed vertices define a consistent classical open string field theory if 
the closed string background satisfies the  classical closed string equations of motion. The inverse assertion does not follow from (\ref{eq:OCHA}). However, it has been shown to be true for infinitesimal closed string deformations in \cite{Sachs open-closed}. More precisely, upon linearizing equation (\ref{eq:OCHA}) in $c\in A_c$ we get
\be\label{eq:OCinf}
N_{cl}(L_{cl}(c))=d_h(N_{cl}(c)) \;\text{.}
\ee
$L_{cl}\in \op{Coder}^{cycl}(SA_c)$ is determined by  $l_{cl}=\pi_1\circ L_{cl}\in\op{Hom}^{cycl}(SA_c)$, the closed string vertices of genus zero (see section \ref{sec:Linfty}). In string field theory the vertex with just one input $(l_{cl})_1$ is the closed string BRST operator 
$Q_c$. Thus equation (\ref{eq:OCinf}) is equivalent to 
\be
N_{cl}(Q_c(c))=d_h(N_{cl}(c))\;\text{,}
\ee
that is $N_{cl}$ induces a chain map from the BRST complex of closed strings to the cyclic Hochschild complex of open string vertices. The cohomology of $Q_c$ (BRST cohomology) defines the space of physical states whereas the cohomology of $d_h$ (cyclic cohomology) characterizes the infinitesimal deformations of the initial open string field theory $M_{cl}$. In \cite{Sachs open-closed} it has been shown that the BRST cohomology of closed strings is indeed isomorphic to the cyclic Hochschild cohomology of open strings. 


\section{Deformations and the quantum open-closed correspondence}\label{sec:defn}
The quantum open-closed homotopy algebra described in the last section is essentially a reformulation of the open-closed BV-equations in terms  of homotopy algebras. However, we can also extract physical insight from this reformulation. The point is that we have the notion of Maurer Cartan elements in homotopy algebras, a concept that is not explicit in the BV formulation. An important property of $IBL_\infty$-morphisms is that they map Maurer Cartan elements into Maurer Cartan elements. Thus a Maurer Cartan element of the closed string loop algebra will in turn define a Maurer Cartan element on the $IBL$-algebra of open string vertices or, in other words, there is a correspondence between certain closed string backgrounds and consistent quantum open string field theory. To make this last statement more precise we will first give a definition of quantum open string field theory and then try to identify corresponding Maurer Cartan elements of the closed string algebra.
 
\subsection{Quantum open string field theory}
To start with we examine the QOCHA in the case where all closed string insertions are set to zero. In equation (\ref{eq:splitopen}) we separated the vertices $\cM$ with open string inputs only from the vertices $\cN$ with both open and closed inputs. Similarly the $IBL_\infty$-morphism separates into
\be\no
\ti{\FF}=e^{\cM}\w \sum_{n=0}^{\infty}\inv{n!}\cN^{\w n}\circ \Delta_n \;\text{.}
\ee
Consider now the defining relation (\ref{eq:QOCHA}) of the QOCHA and set all closed string insertions to zero. We get
\be\label{eq:openloop}
\hbar\;\ti{\FF}\circ D(\omega_c^{-1})=\ti{\LL}_o (e^{\cM}) \;\text{.}
\ee
On the other hand, consistency of the open string field theory implies that $\cM$ has to satisfy the Maurer Cartan equation, that is
\be\label{eq:openSFT}
\ti{\LL}_o (e^{\cM})=0 \;\text{.}
\ee
This is just the quantum BV-equation for open string field theory.  In the classical limit this definition reproduces the known result that the vertices of a classical open string field define an $A_\infty$-algebra \cite{Zwiebach open1,Zwiebach open2}. From (\ref{eq:openloop}) it is then clear that in a trivial closed string background we can have a consistent theory if  $D(\omega_c^{-1})$ is in the kernel of $\ti{\FF}$. 
%
%
%
%
%
As an example we consider Witten's cubic string field theory \cite{Witten cubic,Leclair cubic}.  Cubic string field theory is defined in terms of the BRST operator $Q_o:A_o\to A_o$ and the star product $\ast:A_o\tp A_o \to A_o$. The BRST operator together with the star product define a DGA - a special  case of an $A_\infty$-algebra (see section \ref{sec:Ainfty}). The statement that $Q_o$ and $\ast$ form a DGA solves the constraint imposed by $[\cdot,\cdot]$ whereas the impact of $\delta$ can be summarized as
\be\label{eq:constcubic}
\omega_o(Q_o(e_i),e^i)=0 \mand e_i\ast e^i =0 \;\text{.}
\ee
The first equation in (\ref{eq:constcubic}) is equivalent to demanding that $Q$ is traceless
\be\no
\op{Tr}(Q_o)=0 \;\text{.}
\ee
This is always guaranteed since $Q_o$ is a cohomology operator - $Q_o^2=0$ - and thus the only eigenvalue is zero. 
On the other hand the constraint the star product $\ast$ has to satisfy is more delicate:
\be\no
e_i\ast e^i\stackrel{\mathrm{!}}=0 \;\text{.}
\ee
$e_i\ast e^i$ is precisely the term that arose in the attempt to quantize cubic string field theory \cite{Thorn cubic}. This term is not zero but highly divergent and corresponds to the open string tadpole diagram. The open question is if this divergence can be cured by a suitable regulator, without introducing closed string degrees of freedom explicitly.

\subsection{Quantum open-closed correspondence}
In this section we consider generic closed string backgrounds demanding that they induce consistent quantum open string field theories. In the language of homotopy algebras this property manifests itself in the statement that an $IBL_\infty$-morphism maps Maurer Cartan elements into Maurer Cartan elements. Let us then expand the closed string  Maurer Cartan element, $\mathfrak{c}$ as  
\be\label{gca}
\mathfrak{c}=\sum_{n,g}\hbar^{g+n-1}c^{n,g}\;\,\sep c^{n,g}\in A_c^{\w n}\;\,\sep \LL_c(e^{\mathfrak{c}})=0 \;\text{.}
\ee
Plugging this into equation (\ref{eq:QOCHA}) we get
\be\no
\ti{\LL}_o(\ti{\FF}(e^{\mathfrak{c}}))=0\;\text{.}
\ee
Since $e^{\mathfrak{c}}$ is a constant $IBL_\infty$-morphism and the composition of two morphisms is again a morphisms (see section \ref{sec:IBLinftymorph}), we can conclude that $\ti{\FF}(e^{\mathfrak{c}})$ is a constant morphism and thus
\be\label{eq:quantumOCcorr}
e^{\cM^\prime}\defineL \ti{\FF}(e^{\mathfrak{c}})\sep \ti{\LL}_o(e^{\cM^\prime})=0  \;\text{,}
\ee
where $\cM^\prime=\sum_{g,b}\hbar^{g+b-1}\ti{M^\prime}^{b,g}$ and $\ti{M^\prime}^{b,g}\in \ti{\mathcal{A}}_o^{\w b}$. $\cM^\prime$ represents the open string vertices induced by the closed string Maurer Cartan element via the $IBL_\infty$-morphism. Equation (\ref{eq:quantumOCcorr}) states that these vertices satisfy the requirements of a quantum open string field theory (\ref{eq:openSFT}).
Thus every Maurer Cartan element of the closed string loop algebra defines a quantum open string field theory. 
We give that circumstance a name and call it the quantum open-closed correspondence. To call it a correspondence is maybe a bit misleading. We do not claim that the space of quantum open string field theories is isomorphic to the space of closed string Maurer Cartan elements since we cannot argue that $\ti{\FF}$ is an isomorphism. 
%

An interesting problem is then to find the closed string Maurer Cartan elements or at least to see if  they exist. 
%
The general ansatz for a Maurer Cartan element of an $IBL_\infty$-algebra is given in (\ref{gca}). However, the loop algebra of closed strings is a special case of an $IBL_\infty$-algebra defined by a collection $\sum_g \hbar^g L^g$ of first order coderivations and a second order coderivation $\hbar D(\omega_c^{-1})$. In particular, it defines an $IBL_\infty$-algebra without coderivations of order higher than two. Therefore we claim that a generic Maurer Cartan element of the loop algebra is defined by setting $c^{n,g}=0$ for $n>2$. Explicitly we make the ansatz
\be\no
\mathfrak{c}=c+\hbar g^{-1}\;\text{,}
\ee
where $c\in A_c$ and $g^{-1}\in A_c^{\w2}$. Assume for a moment that $g^{-1}$, considered as a map from $A_c^\ast$ to $A_c$, is invertible and denote its inverse by $g$. Then $g$  defines a metric of degree zero on $A_c$. Let $\{d_i\}$
be a homogeneous basis of $A_c$ and $\{d^i\}$ its dual basis w.r.t. $g$, that is
\be\no
g({}_id, d^j)={}_i\delta^j=g({}^jd, d_i)\;\text{.}
\ee
These two equations are compatible only if we use the sign convention
\be\no
d_i=(-1)^i{}_id \mand d^i={}^id \;\text{.}
\ee
Note that the sign convention for the dual basis of an odd symplectic form is different (see section \ref{sec:loop}). 
With these conventions we can express $g^{-1}$ as
\be\label{eq:invmetric}
g^{-1}=\inv{2}d_i\w{}^id \;\text{.}
\ee
In the following we relax the assumption that $g^{-1}:A_c^\ast \to A_c$ is invertible, but still we can express $g^{-1}$ in the form (\ref{eq:invmetric}) with the corresponding sign convention for $d_i$ and $d^i$.

The Maurer Cartan equation for this particular ansatz reads
\be\label{eq:MCloop}
\bracketi{L_q+\hbar D(\omega_c^{-1})}(e^{c+\hbar g^{-1}})=0 \;\text{,}
\ee
where we abbreviated $L_q=\sum_g \hbar^g L^g$. 

Let us now disentangle this equation and express it in terms of the vertices $l_q=\sum_g\hbar^g l^g=\pi_1\circ L_q$.  A straightforward calculation yields
\be\no
\Delta \bracketii{e^{c+\hbar g^{-1}}}=\sum_{n=0}^\infty\frac{\hbar^{n}}{n!}e^{c+\hbar g^{-1}}\w d_{i_1}\w \dots \w d_{i_n} \tp e^{c+\hbar g^{-1}}\w {}^{i_n}d\w \dots \w {}^{i_1}d \;\text{.}
\ee
Next we plug this into equation (\ref{eq:MCloop}) using (\ref{eq:coderSAex}). Multiplying the resulting equation by $e^{-c-\hbar g^{-1}}$ one obtains
\be\label{eq:MCloopexp}
\sum_{n=0}^\infty \frac{\hbar^n}{n!}\,\bracketi{l_q[c+\hbar g^{-1}]}_n(d_{i_1}\w \dots d_{i_n})\w {}^{i_n}d \w \dots \w {}^{i_1}d \;+\; \hbar \omega_c^{-1}=0\;\text{,}
\ee
where $l_q[c+\hbar g^{-1}]=l_q\circ E(c+\hbar g^{-1})$. 
We proceed by decomposing equation (\ref{eq:MCloopexp}) according to powers in $A_c$, i.e. we project with $\pi_n$ onto $A_c^{\w n}$:
\begin{align}
&\bracketi{l_q[c+\hbar g^{-1}]}_0=0 \label{eq:MCloop1}\\
&\bracketi{l_{q}[c+\hbar g^{-1}]}_1(d_i)\w {}^id \;+\;\omega_c^{-1}=0    \label{eq:MCloop2}\\
&\bracketi{l_{q}[c+\hbar g^{-1}]}_n(d_{i_1}\w\dots\w d_{i_n})\w {}^{i_1}d\w\dots\w {}^{i_n}d=0 \sep n \ge 3\;\text{.}   \label{eq:MCloop3}
\end{align}
Furthermore we can split these equations by comparing coefficients in powers of $\hbar$. Equation (\ref{eq:MCloop1}) gives 
\be\label{eq:MCloopeom}
(l_{cl}[c])_0=\sum_{n=0}^\infty \inv{n!} (l_{cl})_n(c^{\w n})=0 \;\text{.}
\ee
at order $\hbar^{0}$, that is $c$ satisfies the equations of motion of closed string field theory and hence defines a closed string background.
This is what we already had in the classical case. 
The new feature is encoded in equation (\ref{eq:MCloop2}). The $\hbar^{0}$ component of this equation reads
\be\label{eq:MCloopBRST}
(l_{cl}[c])_1(d_i) \w {}^id \;+\; \omega_c^{-1}=0 \;\text{.}
\ee
Note that $(l_{cl}[c])_1$ defines the closed string BRST operator in the new background $c$. We write $(l_{cl}[c])_1=Q_c[c]$. 
Equation (\ref{eq:MCloopBRST}) looks unfamiliar so far, but we can represent it in a more convenient form: First, we make use of the isomorphism
$c_1\w \dots \w c_n\mapsto \inv{n!}\sum_{\sigma\in S_n} c_{\sigma_1}\tp \dots \tp c_{\sigma_n}$ that identifies elements of the symmetric algebra with symmetric tensors.
Equation (\ref{eq:MCloopBRST}) then becomes an equation in $(A_c^{\tp 2})_{sym}$,  the set of second rank symmetric tensors. 
Now act on the second element of this equation with the isomorphism $\omega_c:A_c\to A_c^{\ast}$ and use that $(l_{cl}[c])_1$ is cyclic symmetric w.r.t. $\omega_c$. Following these steps one obtains
\be\label{eq:trivialBRST}
Q_c[c]\circ f+ f \circ Q_c[c]=1\;\text{,}
\ee
where $1$ denotes the identity map on $A_c$ and $f=-g^{-1}\circ \omega_c$. This equation implies that the cohomology of $Q_c[c]$ is trivial. In other words, equation (\ref{eq:trivialBRST}) is saying that $c$ has to be a background where there are no perturbative closed string excitations. This is in agreement with the standard argument that open string field theory is inconsistent due to closed string poles arising at the one loop level. Here, this result arises directly form analyzing the Maurer Cartan element for the closed string $IBL_\infty$-algebra. We should stress, however, that the triviality of the closed string cohomology is neither necessary nor sufficient. It is not sufficient since we have only analyzed the lowest orders in the expansion of the Maurer Cartan equation (\ref{eq:MCloop}) in $d_i$ and $\hbar$. Furthermore, we have not shown that the $IBL_\infty$ map $\FF$ is an isomorphism. Therefore we cannot exclude the existence of Maurer Cartan elements of $\LL_o$ which are not in the image of $\FF$.
%
%
%

To summarize, we found a class of Maurer Cartan elements of the closed string loop algebra involving a background $c\in A_c$ and a linear map $f:A_c\to A_c$ or equivalently an element $g^{-1}\in A_c^{\w 2}$ that can be interpreted, if it is non-degenerate, as the inverse of a metric $g$ on the space of closed string fields. The Maurer Cartan equation implies that $c$ has to be a background that does not admit any physical closed string excitations or, in other words the induced BRST charge $Q_c[c]$ has to have a trivial cohomology. This statement is deduced from equation (\ref{eq:trivialBRST}), which involves the map $f$. There are further implications from the Maurer Cartan equation that are summarized by equation  (\ref{eq:MCloop}). However,  their physical meaning remains unclear and need further investigation.
We chose a special ansatz for the Maurer Cartan elements by setting $c^{n,g}=0$ for $n>2$. However, we find that result conclusion that the closed string background has to have a trivial BRST cohomology persists for a general ansatz for the Maurer Cartan elements.

\section{Outlook}
We showed that $IBL_\infty$ Maurer Cartan elements induce consistent quantum open string field theories. Furhermore, we saw that the Maurer Cartan equation of the closed string loop algebra singles out closed string backgrounds whose associated BRST charge have a trivial cohomology. However, since we have not established that the $IBL_\infty$ map between the closed string loop algebra and the $IBL_\infty$ algebra of open string vertices is an isomorphism the absence of perturbative closed string sates is not proved to be necessary. It would be interesting if such an isomorphism could be established, possibly along the lines of \cite{Sachs open-closed}.  On the other hand, triviality of the closed string cohomology is not sufficient either since there are further implications at higher orders in $\hbar$ whose physical interpretation is not clear yet. Progress in this direction should be useful to classify consistent open string field theories. 

On another front it would be interesting to see how other versions of string field theory such as boundary string field theory \cite{Witten:1992qy,Shatashvili:1993kk} as well as toplogical strings \cite{Witten:1991zz,Bershadsky:1993cx} and refinements thereof \cite{Iqbal:2007ii} fit into the framework of homotopy algebra. Finally, one should expect that there should be a suitable generalization of the homotopy algebras, described here, that encodes the structured of superstring field theory \cite{Berkovits:2000fe}. 
%
%

\vspace{2cm}\noindent {\bf Acknowledgements}: Many parts of this paper rely on a yet unpublished work of Kai Cieliebak et al., and we are very grateful that we had the opportunity to utilize the mathematical framework established therein. We would also like to thank Branislav Jurco and Nicolas Moeller for many fruitful discussions. 
This research was supported in part by DARPA under Grant No.
HR0011-09-1-0015 and by the National Science Foundation under Grant
No. PHY05-51164,  by the DFG Transregional Collaborative Research Centre TRR 33,
the DFG cluster of excellence ``Origin and Structure of the Universe'' as well as the DFG project Ma 2322/3-1.
We would like to thank KITP at UCSB for hospitality during the final stages of this project.

\newpage
\appendix
\section{Symplectic structures in string field theory}\label{app:symplectic}
Here we review the basic ingredients in the formulation of bosonic string field theory \cite{Zwiebach open1,Zwiebach open2,Zwiebach closed,Leclair cubic,Witten cubic}. Strings are described by a conformal field theory on the world sheet, where we denote the spatial resp. time coordinate by $\sigma^1$ resp. $\sigma^2$. This conformal field theory comprises matter and ghosts, where the ghosts arise from gauge fixing the Polyakov action. The space of states $A$ corresponding to that conformal field theory (which is isomorphic to the space of local operators) is the space in which the string fields reside. Furthermore the ghosts endow the vector space $A$ with a $\Z$-grading - the ghost number. In addition we can define an odd symplectic structure $\omega$ on $A$ via the bpz conjugation. This symplectic structure is of outstanding importance for the formulation of string field theory, since the BV operator $\Delta$ and the odd Poisson bracket $(\cdot,\cdot)$ (the two operations that appear in the BV master equation for the string field action $S$)  are constructed with the aid of $\omega$. 

\subsection{Open strings}
The world sheet of an open string is topologically the infinite strip $(0,\pi)\times\R$. By the conformal mapping $z=-e^{-iw}$ ($w=\sigma^1+i\sigma^2$, $(\sigma^1,\sigma^2)\in(0,\pi)\times\R$), the strip is mapped to the upper half plane $H$. The fields living on $H$ can be separated into holomorphic and anti-holomorphic parts, but due to the boundary conditions these two parts combine to a single holomorphic field defined on the whole complex plane $\C$. We expand each field in a Laurant series (mode expansion)
\be
i\pd X^\mu(z)=\sum_{n\in\Z}\frac{\alpha_n^\mu}{z^{n+1}} \ssep c(z)=\sum_{n\in\Z}\frac{c_n}{z^{n-1}} \ssep b(z)=\sum_{n\in\Z}\frac{b_{n}}{z^{n+2}}\;\text{,}
\ee
where the conformal weights are $h_{\pd X}=1$, $h_{c}=-1$, $h_{b}=2$, and the modes satisfy the commutation relations
\be
[\alpha^\mu_m,\alpha^\nu_n]=m\eta^{\mu\nu}\delta_{m+n,0} \ssep \{c_m,b_n\}=\delta_{m+n,0}\;\text{.}
\ee
The space of states $\tilde{A}_o$ is generated by acting with the creation operators on the $SL(2,\R)$ invariant vacuum $|0,k\rangle$, where $k$ denotes the momentum. The grading on $\tilde{A}_o$ is induced by assigning ghost number one to $c$, minus one to $b$ and zero to $X$, i.e. every $c$ mode increases the ghost number by one whereas the $b$ modes decrease the ghost number by one. Utilizing the operator state correspondence, we can identify every state $\Psi\in\tilde{A}_o$ with a local operator $\mathcal{O}_{\Psi}$ and define the bpz inner product by \cite{Leclair cubic}
\be
\left(\Psi_1,\Psi_2\right)_{bpz}\defineL \lim_{z\to 0}\left\langle (I^\ast\mathcal{O}_{\Psi_1})(z)  \mathcal{O}_{\Psi_2}(z) \right\rangle_H\;\text{,}
\ee
where $I(z)=-1/z$, $\langle\dots\rangle_H$ is the correlator on the upper half plane and $I^\ast\mathcal{O}$ denotes the conformal transformation of $\mathcal{O}$ w.r.t. $I$. Since the correlator is $SL(2,\R)$ invariant and $I\in SL(2,\R)$, the bpz inner product is graded symmetric. Note that this correlator is non-vanishing only if it is saturated by three $c$ ghost insertions, i.e. the correlator and consequently the pbz inner product carries ghost number $-3$. The classical string field is an element in $\tilde{A}_o$ of definite ghost number. From the kinetic term of the string field action $S_{kin}=\tinv{2}\bracket{\Psi,Q_o\Psi}_{bpz}$ \cite{Leclair cubic}, where $Q_o$ is the open string BRST charge which carries ghost number one, we can conclude that the classical open string field $\Psi$ must have ghost number one.

Now we would like to identify the bpz inner product with the odd symplectic structure $\omega$, but at first sight this identifications seems to fail since the bpz inner product is graded symmetric rather than graded anti-symmetric. To overcome that discrepancy we shift the degree by one (see section \ref{sec:Ainfty}) which turns an odd graded symmetric inner product into an odd symplectic structure
\be
\omega_o\defineL (\cdot,\cdot)_{bpz}\circ(s\otimes s):A_o\otimes A_o\to\C \;\text{,}
\ee
where $A_o\defineL s^{-1}\tilde{A}_o$. 

To summarize we have an odd symplectic structure $\omega_o$ on $A_o$ of degree $-1$ and the classical open string field is a degree zero element in $A_o$.

\subsection{Closed strings}
The topology of closed strings is that of an infinite cylinder. The conformal mapping $z=e^{-iw}$ maps the cylinder to the complex plane. Now we get twice as many modes as  in the open string since the holomorphic modes are independent of the antiholomorphic ones.
\be
i\pd X^\mu(z)=\sum_{n\in\Z}\frac{\alpha_n^\mu}{z^{n+1}} \ssep c(z)=\sum_{n\in\Z}\frac{c_n}{z^{n-1}} \ssep b(z)=\sum_{n\in\Z}\frac{b_{n}}{z^{n+2}}
\ee 
\be
i\pdo X^\mu(\ov{z})=\sum_{n\in\Z}\frac{\tilde{\alpha}_n^\mu}{\ov{z}^{n+1}} \ssep \tilde{c}(\ov{z})=\sum_{n\in\Z}\frac{\tilde{c}_n}{\ov{z}^{n-1}} \ssep \tilde{b}(\ov{z})=\sum_{n\in\Z}\frac{\tilde{b}_{n}}{\ov{z}^{n+2}}\;\text{.}
\ee
The construction of the vector space $\tilde{A}_c$ is equivalent to that of the open string, except that we constrain the space to the subset of states annihilated by $b_0-\tilde{b}_0$ and furthermore impose the level matching condition \cite{Zwiebach closed}. We assign ghost number one to $c$ and $\tilde{c}$, minus one to $b$ and $\tilde{b}$ and zero to $X$. The correlator on the complex plane $\langle\dots\rangle_\C$ is zero unless we saturate it with three $c$ ghost and three $\tilde{c}$ ghost insertions, i.e. the correlator $\langle\dots\rangle_\C$ has ghost number $-6$. The bpz inner product is defined by \cite{Zwiebach closed}
\be
(\Phi_1,\Phi_2)_{bpz}\defineL \lim_{\abs{z}\to0}\left\langle (I^\ast\mathcal{O}_{\Phi_1})(z,\ov{z}) \mathcal{O}_{\Phi_2}(z,\ov{z})\right\rangle\;\text{,}
\ee
where $\mathcal{O}_{\Phi}$ is again the local operator corresponding to the state $\Phi\in\tilde{A}_c$ and $I(z,\ov{z})=(1/z,1/\ov{z})$.
In contrast to open string field theory the kinetic term of closed string field theory is defined by an additional insertion of $c_0^-=\tinv{2}(c_0-\tilde{c}_0)$, i.e.  $S_{kin}=\tinv{2}(\Psi,c_0^- Q\Psi)_{bpz}$ \cite{Zwiebach closed}. This shows that the ghost number of the classical closed string field $\Phi$ has to be $2$. To unify the presentation we shift the degree by two, such that the classical closed string field is a degree zero element in $A_c\defineL s^{-2}\tilde{A}_c$. The odd symplectic structure of closed string field theory $\omega_c:A_c\otimes A_c\to \C$ is then identified as
\be
\omega_c\defineL \left(\cdot,c_0^{-}\cdot \right)_{bpz}\circ(s^2\otimes s^2)\;\text{.}
\ee
Due to the shift and the $c_0^{-}$ insertion, $\omega_c$ is graded anti-symmetric and has has degree $-1$.

\section{BV master equation and QOCHA}\label{app:BV}
In this section we show that the algebraic relations imposed by the BV master equation in open-closed string field theory are equivalent to the QOCHA. Preliminary we review the BV formalism of open-closed string field theory \cite{Zwiebach open-closed}.

Let $A=\bigoplus_{n}A_n$ be a graded vector space over a field $\F$ endowed with an odd symplectic structure $\omega$ and $\{e_i\}$ be a homogeneous basis of $A$. The dual basis w.r.t. $\omega$ 
is denoted by $\{e^i\}$ 
\be\no
\omega({}_ie,e^j)=\omega({}^je,e_i)={}_i\delta^j \;\text{,}
\ee
where we use again the sign convention ${}_ie=(-1)^ie_i$ and ${}^ie=(-1)^{i+1}e^i$ (see section \ref{sec:loop}). The corresponding bases of forms in $A^\ast$ are denoted by $\{\sigma_i\}$ and $\{\sigma^i\}$, i.e.
\be\no
{}_i\sigma(e^j)={}_i\delta^j = {}^j\sigma(e_i)\;\text{.}
\ee
Consistency of these two equations requires the sign convention $\sigma^i={}^i\sigma$ and $\sigma_i={}_i\sigma$. 
We can consider the vector space $A$ as a supermanifold. The points in this supermanifold are vectors $c\in A$ and expressed in components $c=c^i{}_ie=c_i{}^ie=e_i{}^ic=e^i{}_ic$. The tangent space of that manifold is spanned by the collection of derivatives w.r.t. the components of $c$. We distinguish between left and right derivatives. A left resp. right derivative acts from the left resp. right and is labelled by an arrow $\rightharpoonup$ resp. $\leftharpoonup$.
We define
\begin{align}\no
&{}_i\pd\defineL\overset{\rightharpoonup}{\partial}_{c^i} \sep {}^i\pd\defineL \overset{\rightharpoonup}{\partial}_{c_i}\;\text{,}\\\no
&\pd_i\defineL \overset{\leftharpoonup}{\partial}_{{}^ic} \sep \pd^i\defineL \overset{\leftharpoonup}{\partial}_{{}_ic}\;\text{,}
\end{align}
and the differential of a function $f\in C^\infty(A)$ is defined by
\be\no
df=\sigma^i\,{}_i\pd f=\sigma_i\, {}^i\pd f = f\pd_i \, {}^i\sigma=f\pd^i \,{}_i\sigma \;\text{.}
\ee
With this convention we have for example ${}_i\pd c={}_ie$.

To every function $f\in C^\infty(A)$ we can assign a Hamiltonian vector field $X_f\in \text{Vect}(A)$ by
\be
df=-i_{X_f}\omega \;\text{,}
\ee
where $i_X$ denotes the interior product, i.e. the contraction w.r.t. to the vector field $X$.
The odd Poisson bracket (antibracket) $(\cdot,\cdot)$ is then defined by \cite{Schwarz bv}
\be\label{eq:oddpoisson}
(f,g)=X_f(g) \;\text{,}
\ee
for $f,g \in C^\infty(A)$. The BV operator $\Delta$ is defined by \cite{Schwarz bv}
\be\label{eq:bvoperator}
\Delta f= \inv{2}\op{div}X_f \;\text{,}
\ee
and squares to zero $\Delta^2=0$ since $\omega$ has degree $-1$. Here we suppress superscript $BV$ for the BV operator  since it cannot be confused with the comultiplications in the present context.
In components we get 
\be\no
(f,g)=(-1)^{i}f\pd_i\,{}^i\pd g \mand \Delta f = \inv{2}{}_i\pd\,{}^i\pd f \;\text{.}
\ee
Equivalently the odd Poisson bracket can be defined via the BV operator 
\be
(f,g)=(-1)^f\bracketii{\Delta(fg)-\Delta(f)g-(-1)^ff\Delta(g)}\;\text{,}
\ee
i.e. the odd Poisson bracket is the deviation of $\Delta$ being a derivation. The BV operator is indeed a second order derivation \cite{Getzler bv}, that is
\begin{align}\no
&\Delta(fgh)-\Delta(fg)h-(-1)^{f(g+h)}\Delta(gh)f-(-1)^{h(f+g)}\Delta(hf)g\\\no
&+\Delta(f)gh+(-1)^{f(g+h)}\Delta(g)hf+(-1)^{h(f+g)}\Delta(h)fg=0\;\text{.}
\end{align}
Furthermore the following identities hold \cite{Getzler bv}:
\begin{align}\no
\Delta(f,g)&=(\Delta f,g)+(-1)^{f+1}(f,\Delta g)\\\no
0&=(-1)^{(f+1)(g+1)}(f,(g,h))+(-1)^{(g+1)(f+1)}(g,(h,f))+(-1)^{(h+1)(g+1)}(h,(g,f))\\\no
(f,gh)&=(f,g)h+(-1)^{(f+1)g}g(f,h)
\end{align}
The first is saying that $\Delta$ is a derivation over $(\cdot,\cdot)$, the second is the Jacobi identity for $(\cdot,\cdot)$ and the third is saying that $(f,\cdot)$ is a derivation of degree $|f|+1$ on the space of functions.

In open-closed string field theory the vector space is the direct sum $A=A_o\oplus A_c$ and the symplectic structure is $\omega=\omega_o\oplus\omega_c$. Hence the BV operator and the odd Poisson bracket also split into open and closed parts:
\be\no
\Delta=\Delta_o + \Delta_c \sep (\cdot,\cdot)=(\cdot,\cdot)_o+(\cdot,\cdot)_c\;\text{.}
\ee
The quantum BV master equation reads 
\be\no
\hbar\Delta S+\inv{2}(S,S)=0\;\text{,}
\ee
where 
\be\no
S=\sum_{g=0}^\infty \hbar^{2g-1}(\omega_c\circ l^g)(e^{\hbar^{1/2}c})+ \sum_{b=1}^\infty\sum_{g=0}^\infty \inv{b!}\hbar^{2g+b-1}(\tilde{\omega}^{\tp b}_o\circ \tilde{f}^{b,g})(e^{\hbar^{1/2}c};\bar{e}^a,\dots,\bar{e}^a)\;\text{,}
\ee
is the BV action of equation (\ref{eq:BVaction}). Before we consider the general case, let us restrict to open string field theory in the classical limit. In this case the action reads
\be
S_{o,cl}=\sum_{n=1}^\infty \inv{n+1}\omega_o(m_n(a^{\tp n}),a)\;\text{.}
\ee 
The classical BV equation of open string field theory is
\begin{align}\label{eq:BVopencl}
(S_{o,cl},S_{o,cl})_o&=(-1)^{i}\sum_{n_1=1}^\infty \sum_{n_2=1}^\infty \inv{n_1+1}\inv{n_2+1}\omega_o\bracketi{m_{n_1}(a^{\tp n_1}),a}\,\pd_i\; {}^i\pd\, \omega_o\bracketi{m_{n_2}(a^{\tp n_2}),a}\\\no
&=(-1)^{i}\sum_{n_1=1}^\infty \sum_{n_2=1}^\infty \omega_o\bracketi{m_{n_1}(a^{\tp n_1}),e_i}\, \omega_o\bracketi{m_{n_2}(a^{\tp n_2}),{}^ie}\\\no
&=\omega_o\bracketi{m(e^a),e_i}\,\omega_o\bracketi{{}^ie,m(e^a)}=\omega_o\bracketi{m(e^a),e_i}\,{}^i\sigma\bracketi{m(e^a)}\phantom{\sum_{n_1=1}^\infty}\\\no
&=\omega_o\bracketi{m(e^a),m(e^a)}=\sum_{n=1}^\infty \frac{2}{n+2}\sum_{i+j+k=n}\omega_o\bracketii{m_{i+k+1}\bracketi{a^{\tp i}\tp m_j(a^{\tp j})\tp a^{\tp k}},a}\\\no
&=\sum_{n=1}^\infty \frac{2}{n+2}\omega_o\bracketi{\pi_1\circ M^2(a^{\tp n}),a}=0\;\text{.}
\end{align}
All we had to use is cyclicity of $m_n$ and $e_i\tp {}^i\sigma=1$, where $1$ denotes the identity map on $A_o$. $M\in \op{Coder}^{cycl}(TA_o)$ is the coderivation corresponding to $m\in\op{Hom}^{cylc}(TA_o,A_o)$ and equation (\ref{eq:BVopencl}) is equivalent to $M^2=0$, the well known statement that the vertices of a classical open string field theory define an $A_\infty$-algebra \cite{Zwiebach open1}. Schematically we write
\be\no
(S_{o,cl},S_{o,cl})_o\sim \omega_o\bracketii{m\bracketi{e^a\tp m(e^a)\tp e^a},a}\;\text{,}
\ee
i.e. $\sim$ indicates that we will ignore the precise coefficients. 

In order to keep the presentation clear, we will use this notation for the treatment of the quantum BV action of open and closed strings. Furthermore we abbreviate $c^\prime=\hbar^{1/2}c$.
We just collect the results here since the calculations are quite similar to that in (\ref{eq:BVopencl}). From the open string BV operator we get
\begin{align}\label{eq:openBV}
\hbar\Delta_o S \sim& \phantom{+}\sum_{b,g}\hbar^{2g+b}\sum_{k=1}^b\ti{\omega}_o^{\tp b}\circ\ti{f}^{b,g}(e^\cp;e^a,\dots,\underbrace{e_i\tp e^a\tp e^i\tp e^a}_{\text{$k$-th boundary}},\dots,e^a)\\\no
&+\sum_{b,g}\hbar^{2g+b}\sum_{k\neq l}\ti{\omega}_o^{\tp b}\circ\ti{f}^{b,g}(e^\cp;e^a,\dots,\underbrace{e_i\tp e^a}_{\text{$k$-th bdry}},\dots,\underbrace{e^i\tp e^a}_{\text{$l$-th bdry}},\dots,e^a)  \;\text{.} 
\end{align}
The first term in (\ref{eq:openBV}) translated into homotopy language is equivalent to
\be\no
\sum_{b,g}\hbar^{2g+b}\;\hati{\delta}\bracketi{\ti{F}^{b,g}}\;\text{,}
\ee
whereas the second term is equivalent to
\be\no
\sum_{b,g}\hbar^{2g+b}\;\hati{[\cdot,\cdot]}\bracketi{\ti{F}^{b,g}}\;\text{.}
\ee
Here we see that $\Delta_o$ partly translates into $\hati{[\cdot,\cdot]}$ as anticipated in section \ref{sec:QOCHA}.
The closed string BV operator contributes
\begin{align}\label{eq:closedBV}
\hbar\Delta_c S \sim& \phantom{+}\sum_{g}\hbar^{2g+1}\omega_c\circ l^g(\omega_c^{-1}\w e^\cp)\\\no
&+\sum_{b,g}\hbar^{2g+b+1}\ti{\omega}_o^{\tp b}\circ \ti{f}^{b,g}(\omega_c^{-1}\w e^\cp;e^a,\dots,e^a)
  \;\text{.} 
\end{align}
The second equation in (\ref{eq:closedBV}) is equivalent to
\be\no
\sum_{b,g}\hbar^{2g+b+1}\ti{F}^{b,g}\circ D(\omega_c^{-1})\;\text{.}
\ee
Next consider the open string Poisson bracket
\begin{align}\label{eq:openPoisson}
(S,S)_o \sim& \phantom{+}\sum_{g_1,g_2 \atop b_1,b_2}\hbar^{2(g_1+g_2)+b_1+b_2-2 }\sum_{k,l=0}^b \;\ti{\omega}_o\circ \ti{f}^{b_1,g_1}(e^\cp;e^a,\dots,\underbrace{e_i\tp e^a}_{\text{$k$-th bdry}},\dots,e^a)\\\no
&\hspace{4.75cm}\cdot\,\ti{\omega}_o\circ \ti{f}^{b_2,g_2}(e^\cp;e^a,\dots,\underbrace{e_i\tp e^a}_{\text{$l$-th bdry}},\dots,e^a)\;\text{.} 
\end{align}
Equation (\ref{eq:openPoisson}) is equivalent to 
\be\label{eq:openPoissonco}
\sum_{g_1,g_2 \atop b_1,b_2}\hbar^{2(g_1+g_2)+b_1+b_2-2 } \,\inv{2}\hati{[\cdot,\cdot]}^\prime\bracketi{\ti{F}^{b_1,g_1}\w\ti{F}^{b_2,g_2}}\circ\Delta\;\text{,}
\ee
where the prime $\prime$ indicates that the first resp. second input must be out of $\ti{F}^{b_1,g_1}$ resp. $\ti{F}^{b_2,g_2}$. 
If we express (\ref{eq:openPoissonco}) in terms of the unrestricted $\hati{[\cdot,\cdot]}$, we have to compensate by subtracting twice the part where $\hati{[\cdot,\cdot]}$ acts on only one of the $\ti{F}$'s, i.e. 
\begin{align}\no
&\sum_{g_1,g_2 \atop b_1,b_2}\hbar^{2(g_1+g_2)+b_1+b_2-2 } \,\inv{2}\hati{[\cdot,\cdot]}^\prime\bracketi{\ti{F}^{b_1,g_1}\w\ti{F}^{b_2,g_2}}\circ\Delta\\\no
&=\sum_{g_1,g_2 \atop b_1,b_2}\hbar^{2(g_1+g_2)+b_1+b_2-2 }\bracket{\inv{2}\hati{[\cdot,\cdot]}\bracketi{\ti{F}^{b_1,g_1}\w\ti{F}^{b_2,g_2}} \,-\,\bracketi{\hati{[\cdot,\cdot]}\ti{F}^{b_1,g_1}}\w\ti{F}^{b_2,g_2}    }\circ\Delta 
\end{align}

Finally the closed string Poisson bracket yields
\begin{align}\label{eq:closedPoisson}
(S,S)_c \sim& \phantom{+}\sum_{g_1,g_2}\hbar^{2g_1+2g_2-1}\;\omega_c\bracketii{l^{g_1}\bracketi{l^{g_2}(e^\cp)\w e^\cp},\cp}\\\no
&+ 2\sum_{g_1,g_2, b}\hbar^{2(g_1+g_2)+b-1}\;\ti{\omega}_o^{\tp b}\circ\ti{f}^{b,g_1}\bracketi{l^{g_2}(e^\cp)\w e^\cp;e^a,\dots,e^a}\\\no
&+ \sum_{g_1,g_2 \atop b_1,b_2}\hbar^{2(g_1+g_2)+b_1+b_2-1}\;\ti{\omega}_o^{\tp b_1}\circ \ti{f}^{b_1,g_1}\bracketi{e_i\w e^\cp;e^a,\dots,e^a}\\\no
&\hspace{4cm}\cdot\,\ti{\omega}_o^{\tp b_2}\circ \ti{f}^{b_2,g_2}\bracketi{e^i\w e^\cp;e^a,\dots,e^a}\;\text{.}
\end{align}
The second term in (\ref{eq:closedPoisson}) is equivalent to
\be\no
\sum_{g_1,g_2, b}\hbar^{2(g_1+g_2)+b-1}\ti{F}^{b,g_1}\circ L^{g_2}\;\text{,}
\ee 
while the third term is equivalent to
\be\no
\sum_{g_1,g_2 \atop b_1,b_2}\hbar^{2(g_1+g_2)+b_1+b_2-1} \bracketi{\ti{F}^{b_1,g_1}\circ D(e_i)\w \ti{F}^{b_2,g_2}\circ D(e^i)}\circ\Delta \;\text{.}
\ee
We see that the third term is associated with $D(\omega_c^{-1})$, that is $(\cdot,\cdot)_c$ plays partly the role of $D(\omega_c^{-1})$ as we pointed out in \ref{sec:QOCHA}.
The fact that second order derivations in the BV formalism translate into not just second order but also first oder coderivations in the homotopy language and vice versa, is actually the reason why the powers in $\hbar$ in the BV formalism ($\hbar^{2g+b+n/2-1}$) differ from that in the homotopy language ($\hbar^{g+b-1}$).

Let us collect the individual terms now.
First consider the terms with closed strings only. By comparing coefficients in $\hbar$, we recover the loop algebra of closed strings:
\be\no
\sum_{g_1+g_2=g \atop i_1+i_2=n}\sideset{}{^\prime}\sum_\sigma l^{g_1}_{i_1+1}\circ(l^{g_2}_{i_2}\w 1^{\w i_1})\circ \sigma \,+\, l^{g-1}_{n+2}(\omega_c^{-1}\w 1^{\w n})=0
\ee
Finally turn to the parts with open and closed strings. First project onto $\ti{\mathcal{A}}_o^{\w b}$, i.e. terms with a definite number of boundaries, and then compare coefficients in $\hbar$. Following that procedure we precisely obtain the QOCHA:
\begin{align}\no
&\cF\circ\LL_c+\frac{\hbar}{2} \bracketi{\cF\circ D(e_i)\w\cF\circ D(e^i)}\circ\Delta \\\no
 &=\ti{\LL}_o\circ \cF + \inv{2}\hati{[\cdot,\cdot]}\circ\bracketi{\cF\w \cF}\circ \Delta -\bracket{(\hati{[\cdot,\cdot]}\circ\cF)\w \cF}\circ \Delta\;\text{.}
\end{align}


\end{document}